\documentclass[reprint,superscriptaddress,footinbib, amsmath,amssymb,aps,pre,onecolumn]{revtex4-2}
\usepackage{xr,mathtools}
\usepackage{hyperref}
\usepackage{gibbon}
\usepackage{lmodern}
\usepackage[dvipsnames]{xcolor}
\usepackage{dcolumn}
\usepackage{tabularx}
\usepackage[framemethod=tikz]{mdframed} 
\usepackage{listings,xcolor}
\usepackage{empheq}
\usepackage{ulem}[normalem]

\newcommand{\noop}[1]{}
\newcommand{\add}[1]{\color{black}{#1 }\color{black}}

\newcommand{\moveI}[1]{\color{black}{#1 }\color{black}}

\begin{document}

\title{{Ideal incompressible axisymmetric MHD: Uncovering finite-time singularities}}
\author{Sai Swetha Venkata Kolluru}
 \email{saik@iisc.ac.in}
\author{Rahul Pandit}
\email{rahul@iisc.ac.in}
\affiliation{Centre for Condensed Matter Theory, Department of Physics, Indian Institute of Science, Bengaluru, India - 560012}
\date{\today}

\begin{abstract}
We provide compelling numerical evidence for the development of (potential) finite-time singularities in the three-dimensional (3D) axisymmetric, ideal, incompressible magnetohydrodynamic (IMHD) equations, in a wall-bounded cylindrical domain, starting from smooth initial data, for the velocity and magnetic fields. 
We demonstrate that the nature of the singularity depends crucially on the relative strength $\mathcal{C}$ of the velocity and magnetic fields at the time of initialisation: (i) if $\mathcal{C} < 1$, then the swirl components, at the wall, evolve towards square profiles that lead  to the intensification of shear at the meridional plane $(r = 1, z = L/2)$ and the development of a finite-time singularity; (ii) if $\mathcal{C} = 1$, there is no temporal evolution; (iii) if $\mathcal{C} > 1$, then the swirl components, at the wall, evolve towards \noop{structures with} a cusp-type singularity. By examining the spatiotemporal evolution of the pressure, we obtain \noop{deep} insights into the development of these singularities .
\end{abstract}
\maketitle

{Among the many interesting problems posed by the partial differential equations (PDEs) of fluid dynamics, perhaps the most challenging is
proving the global regularity of \add{their} solutions. For the three-dimensional (3D) Navier--Stokes equation (NSE), which governs the motion of a Newtonian fluid~\cite{doering1995applied,galdi2000introduction,foias2001navier,robinson2016three,lemarie2018navier,robinson2020navier}, the global-regularity problem is one of the Clay Millennium Prize Problems~\cite{carlson2006millennium}~\cite{Noteme}. A proof of global regularity (or lack thereof) should show that, starting with smooth initial data for the hydrodynamical fields, solutions of the PDE do not develop (or develop) singularities in finite time. For the two-dimensional (2D) NSE and Euler equation, global regularity has been proved for analytic initial data~\cite{yudovich1963non}. 
By contrast, the proofs of the global regularity of solutions of the 3D NSE and its inviscid forerunner, the 3D Euler equation, continue to be open problems. Furthermore, the breakdown of the regularity of solutions of the 3D Euler equation has been conjectured to be important for understanding anomalous dissipation for turbulent solutions of the 3D NSE, in the limit of vanishing viscosity~\cite{onsager,eyink2006onsager}.}\\

{The global regularity problem remains open in both 2D and 3D~\cite{paicu2021global} even for the ideal, incompressible MHD (IMHD) equations, which govern the dynamics of an incompressible, inviscid conducting fluid with zero magnetic diffusivity and are of relevance in astrophysics, space physics, geophysics, nuclear fusion, and liquid-metal flows~\cite{biskamp2003magnetohydrodynamic,choudhuri1998physics,goedbloed2004principles,freidberg2014ideal,galtier2016introduction,davidson2017introduction,gurnett2017introduction,goedbloed2019magnetohydrodynamics}. 
The existence (or nonexistence) of a finite-time singularity (FTS) in \add{the solutions of} the IMHD equations is of \noop{central} importance not only from \add{a mathematical standpoint} \noop{the  standpoint of mathematics}, but also from the point-of-view of physics~\cite{brachet2013ideal}.
Numerical studies of 3D incompressible MHD turbulence have shown that, at high Reynolds numbers and high magnetic Reynolds numbers, the rate of energy dissipation reaches a positive constant value~\cite{dallas2014signature,mininni2009finite,linkmann2015nonuniversality}; this phenomenon is reminiscent of anomalous dissipation at high Reynolds numbers in 3D NS turbulence. \add{Consequently,} it has been conjectured that the breakdown of global regularity for the 3D IMHD equations might be related to anomalous dissipation in MHD turbulence~\cite{caflisch1997remarks}.}

{Numerical investigations of both the 3D incompressible Euler~\cite{gibbon2008euler,hou2009blow} and the 3D IMHD equations~\cite{kerr1999evidence,grauer2000current,brachet2013ideal}, in spatially periodic domains, have not yielded conclusive evidence for or against an FTS. 
By contrast, there is growing numerical evidence for a potential FTS in the 3D-axisymmetric incompressible Euler (3DAE) equations, in a wall-bounded cylindrical domain.
This was first \add{reported in the} \noop{explored in the high-order Galerkin and finite-difference} adaptive-mesh calculation of Ref.~\cite{luo2014potentially} which reported the development of a ring-like singularity on the wall of the cylinder in the meridional plane at an estimated singularity time $t_* \simeq 0.0035056$. This potential FTS was re-examined and confirmed using Fourier-Chebyshev pseudospectral~\cite{barkley,kolluru2022insights} and Cauchy-Lagrange~\cite{hertel2022cauchy} methods. The role of pressure in the development of this FTS was elucidated in Ref.~\cite{barkley}.}

\add{In this paper,} we provide compelling numerical evidence for a similar FTS for the 3D-axisymmetric IMHD equations in a wall-bounded domain, with smooth initial data for both velocity and magnetic fields [akin to the initial condition in Refs.~\cite{luo2014potentially,kolluru2022insights} for the 3DAE]. Furthermore, we uncover a new type of FTS by tuning the ratio $\mathcal{C}$ of the magnitudes of the initial magnetic and velocity fields.  If  $\mathcal{C} < 1$, the evolution of the FTS is akin to that in the 3DAE~\cite{luo2014potentially,kolluru2022insights,hertel2022cauchy}. 
In contrast, if $\mathcal{C} > 1$, then the swirl-components of the fields develop cusp-like structures at the wall that lead, in turn, to a potential FTS.
{This novel cusp-type singularity has not been seen for any axisymmetric hydrodynamical partial differential equation}~\cite{Note2}
For both cases $\mathcal{C} < 1$ and $\mathcal{C} > 1$, the vorticity and current fields show rapid growth.

\moveI{The evolution of an incompressible, inviscid, perfectly conducting fluid \noop{of unit density} is governed by the incompressible, ideal MHD equations (IMHD) \noop{given} below \noop{, where $\bu$ represents the velocity of the fluid and $\bdb$ its intrinsic magnetic field}:
\begin{subequations}
\begin{align}
\partial_t \bu + \bu\cdot \nabla \bu & = {- \nabla p }+ \bj \times \bdb\,; \label{eq:imhd_u} \\
\partial_t \bdb & = \nabla \times ( \bu \times \bdb )\,. \label{eq:imhd_b}
\end{align}
\label{eq:imhd_full}
\end{subequations}}
\add{In cylindrical coordinates, the velocity field is written as $\bu (r,z,t)= (u^r(r,z,t), u^{\theta}(r,z,t), u^z(r,z,t))$. Here, $\nabla \cdot \bu = 0 $ and $\nabla \cdot \bdb = 0 $.;
there is a similar expression for the magnetic field $\bdb$; the superscripts $r$, $\theta$, and $z$ denote the radial, swirl, and axial components of the fields.} \noop{[replaces]The 3D-axisymmetric IMHD equations~\eqref{eq:mhd_pressure}-~\eqref{eq:main_rz} are conveniently written in cylindrical coordinates with the velocity field $\bu (r,z,t)=u^r(r,z,t) \ \hat{e}_r+u^{\theta}(r,z,t) \ \hat{e}_{\theta}+u^z(r,z,t) \ \hat{e}_z$, where $\hat{e}_r, \,\hat{e}_{\theta}$ and $\hat{e}_z$ are the basis vectors;
there is a similar expression for the magnetic field $\bdb$; the superscripts $r$, $\theta$, and $z$ denote the radial, swirl, and axial components of the fields.}\\ 
\moveI{We solve the \add{IMHD equations} in an axisymmetric cylindrical domain: in the $z$-direction, we use periodic boundary conditions for all fields and in the $r$ direction, we impose no-flow condition for $\bu$, and the perfectly conducting condition for $\bdb$, at the wall, i.e., $u_r(r=1,z,t)=b_r (r=1,z,t)=0$; (ii) at the axis, we use the pole conditions for all fields, namely, that their radial derivatives vanish at $r=0$~\cite{houli,kolluru2022insights}.}
\noop{We solve these equations, with the boundary conditions \noop{(BC)} and initial conditions given in Eqs.~\eqref{eq:bc}-~\eqref{eq:ic}, by developing a Fourier-Chebyshev pseudospectral method [Eq.~\eqref{eq:FC}].} \\
\moveI{We initialize the velocity field with $u^{\theta}(r,z,t=0)/r = 100 e^{-30(1-r^2)^4} \sin{\left(\tfrac{2\pi}{L}z \right)}$ as in Refs.~\cite{luo2014potentially,kolluru2022insights}; and the initial magnetic field $b^{\theta}(r,z,t=0) = \mathcal{C} u^{\theta}(r,z,t=0)$ is chosen to be a multiple of the initial velocity where $\mathcal{C}$ controls the strength of $b^{\theta}$ relative to $u^{\theta}$, at the initial time.}

\begin{figure}[htbp]
\centering
    \raisebox{0.07\height}{
    \begin{tikzpicture}
    \draw (0,0) node[inner sep=0]{\includegraphics[height=0.28 \linewidth]{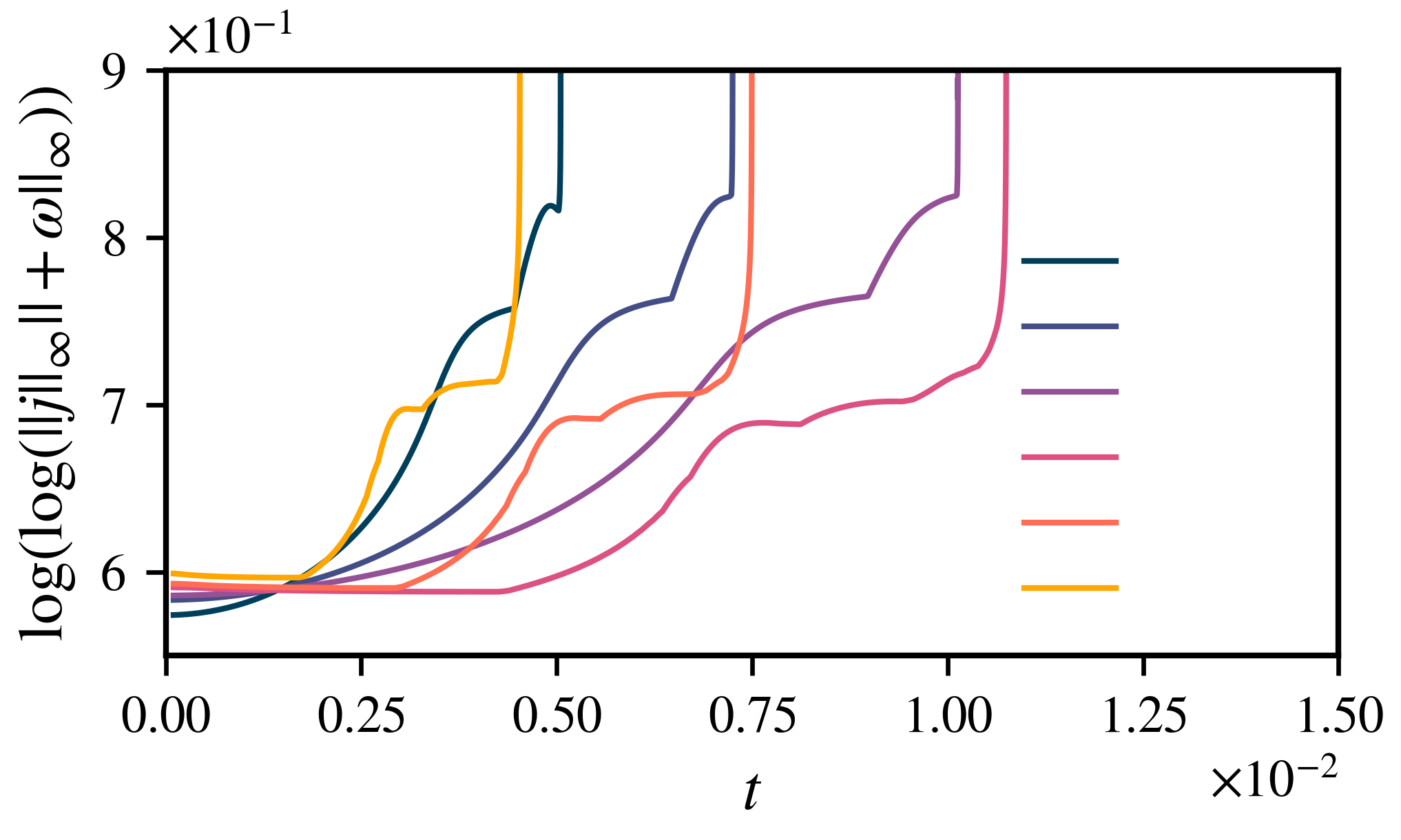}};
    \draw (-2.5,1.6) node {(a)};
    \draw (2.75,0.85) node {$\scriptstyle \mathcal{C}=0.50$};
    \draw (2.75,0.52) node {$\scriptstyle \mathcal{C}=0.80$};
    \draw (2.75,0.16) node {$\scriptstyle \mathcal{C}=0.90$};
    \draw (2.75,-0.19) node {$\scriptstyle \mathcal{C}=1.10$};
    \draw (2.75,-0.56) node {$\scriptstyle \mathcal{C}=1.20$};
    \draw (2.75,-0.92) node {$\scriptstyle \mathcal{C}=1.50$};
    \end{tikzpicture}}
    \hspace{1em}
    \begin{tikzpicture}
    \draw (0,0) node[inner sep=0]{\includegraphics[height=0.28\linewidth]{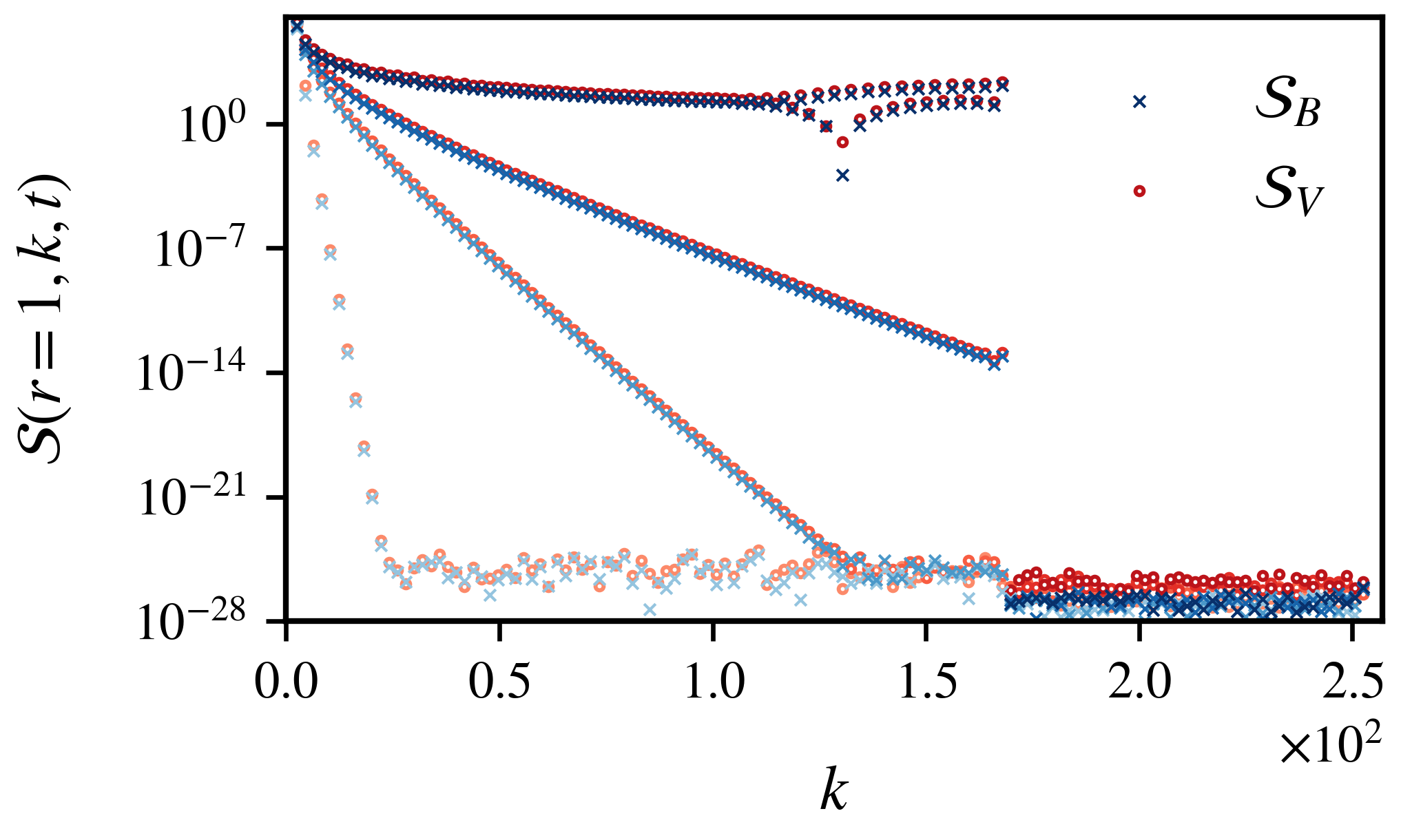}};
    \draw (-1.9,-0.8) node {(b)};
    \end{tikzpicture}
    \caption{(a) Plot vs. $t$ of the CKS criterion $\log(\log(||\bj ||_{\infty} + ||\bom ||_{\infty}))$ for different values of $\mathcal{C}$ given in the legend. We see rapid growth of the CKS integrand in time, which occurs earlier for values of $\mathcal{C}$ far from $\mathcal{C}=1$. (b) Semi-log plots versus $k$ of the Fourier energy spectra $\mathcal{S}_V(r=1,k,t)$ (red) and $\mathcal{S}_B(r=1,k,t)$ (blue) at the wall (at $r=1$) for four different times $t=\{ 0.0009, 0.0034, 0.0041, 0.0062\}$. For early times, these spectra  (shown in faded colours) have exponential tails, with large negative slopes, so our system exhibits spectral convergence. At later times, the spectra flatten out and begin to thermalise (see text); the flow is initiated with the representative value $\mathcal{C}=0.80$ and a resolution of $(N=256, M=512)$. }
    \label{fig:80_spectra}
\end{figure}
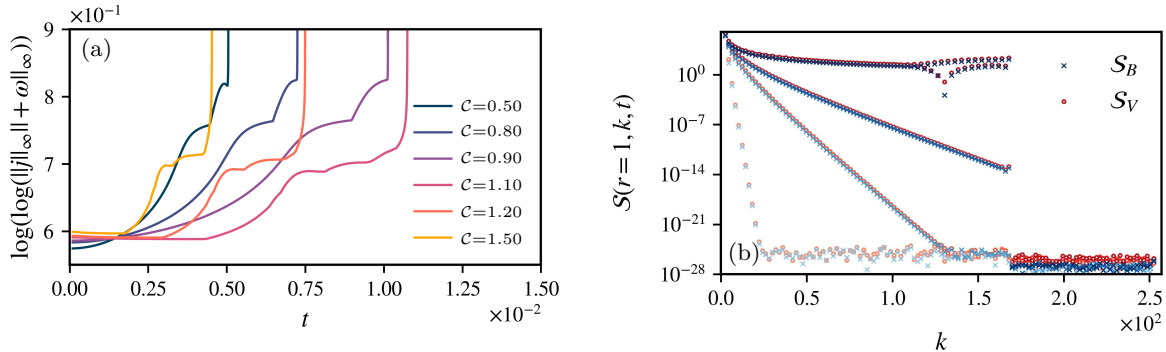

We \add{first} analyse \noop{the development of FTS} \add{the IMHD flows evolved from the aforementioned initial data} using the Caflisch-Klapper-Steele (CKS) criterion \add{\cite{caflisch1997remarks} in Fig.~\ref{fig:80_spectra}(a).}
\noop{~\eqref{eq:cks}}  \noop{and then the analyticity-strip method [Eq.~\eqref{eq:analyticity-strip}], which employs velocity and magnetic-field spectra [Eqs.~\eqref{eq:Spectra_V}-~\eqref{eq:Spectra}] and well-suited for our pseudospectral study. We validate our numerical method by examining the temporal evolution of the conserved quantities, namely, the total (kinetic + magnetic) energy $E_T$, magnetic helicity $H_M$, and the cross helicity $H_C$ [Eqs.~\eqref{eq:invariant_et}-~\eqref{eq:invariant_hm} in Section~\ref{subsec:conserved} and Fig. F1 in the Supplemental Information].}
\noop{The Beale-Kato-Majda (BKM) criterion~\cite{beale1984remarks}, which was developed to look for FTSs for the 3D Euler equations~\cite{beale1984remarks}, was generalised to examine 
FTSs in ideal MHD flows by Caflisch, Klapper, and Steele~\cite{caflisch1997remarks}. 
We first employ the Caflisch-Klapper-Steele (CKS) criterion [Section~\ref{subsec:Methods_detect}] to check for the 
development of a potential FTS in the 3D-axisymmetric IMHD equations: }The rapid growth of $||\bj||_{\infty}+||\bom||_{\infty}$, the 
integrand in the CKS criterion~\eqref{eq:cks}, indicates the generation of \noop{such} FTSs  for various values of $\mathcal{C}$. \noop{ [Fig. F1 (d) in the Supplemental Information].}
\add{However, since we employ a fixed-grid spectral scheme,} we \add{do not} \noop{cannot} rely completely on the CKS criterion for singularity detection. \noop{, in fixed-grid spectral simulations, so} 
We now carry out a detailed study of these potential FTSs using the analyticity-strip method \add{of Ref.\cite{sulem} which is especially well-suited for spectral schemes}. 

{In order to track the nearest complex-space singularities using the analyticity-strip method, we consider the Fourier spectra $\mathcal{S}_V(r=1,k,t)$ and $\mathcal{S}_B(r=1,k,t)$ of the velocity and magnetic fields \noop{[Eqs.~\eqref{eq:Spectra} in Section~\ref{subsec:Methods_detect}]} computed at the wall ($r=1$), where we expect the potential FTS to precipitate, \add{in Fig.~\ref{fig:80_spectra}(b)}.

\noop{ We show semi-log plots of these spectra in red and blue, respectively, in Fig.~\ref{fig:80_spectra}, for representative value of the time $t$, increasing from $t= 0.0009$ to $ t = 0.0062$, for IMHD flows with \noop{the BC~\eqref{eq:bc} and} the initial condition~\eqref{eq:ic}; for purposes of illustration we present spectra for $\mathcal{C} = 0.8$ [i.e., $C=80$].}
At large-$k$, the spectra exhibit exponentially decaying tails, so we use the asymptotic relation~\eqref{eq:analyticity-strip} to extract the slope $\delta_T(t)$ of the of the semi-log plot in Fig.~\ref{fig:80_spectra}(b),
which yields our estimate for the distance of the nearest complex singularity from the real line, at time $t$~\cite{Note3}. As time progresses, we find that $\delta_T \rightarrow 0$, i.e., the complex singularity approaches the real line,
and a potential FTS develops.} 

\begin{figure}[htbp]
    \centering
    \begin{tikzpicture}
    \draw (0,0) node[inner sep=0]{\includegraphics[height=0.32\linewidth]{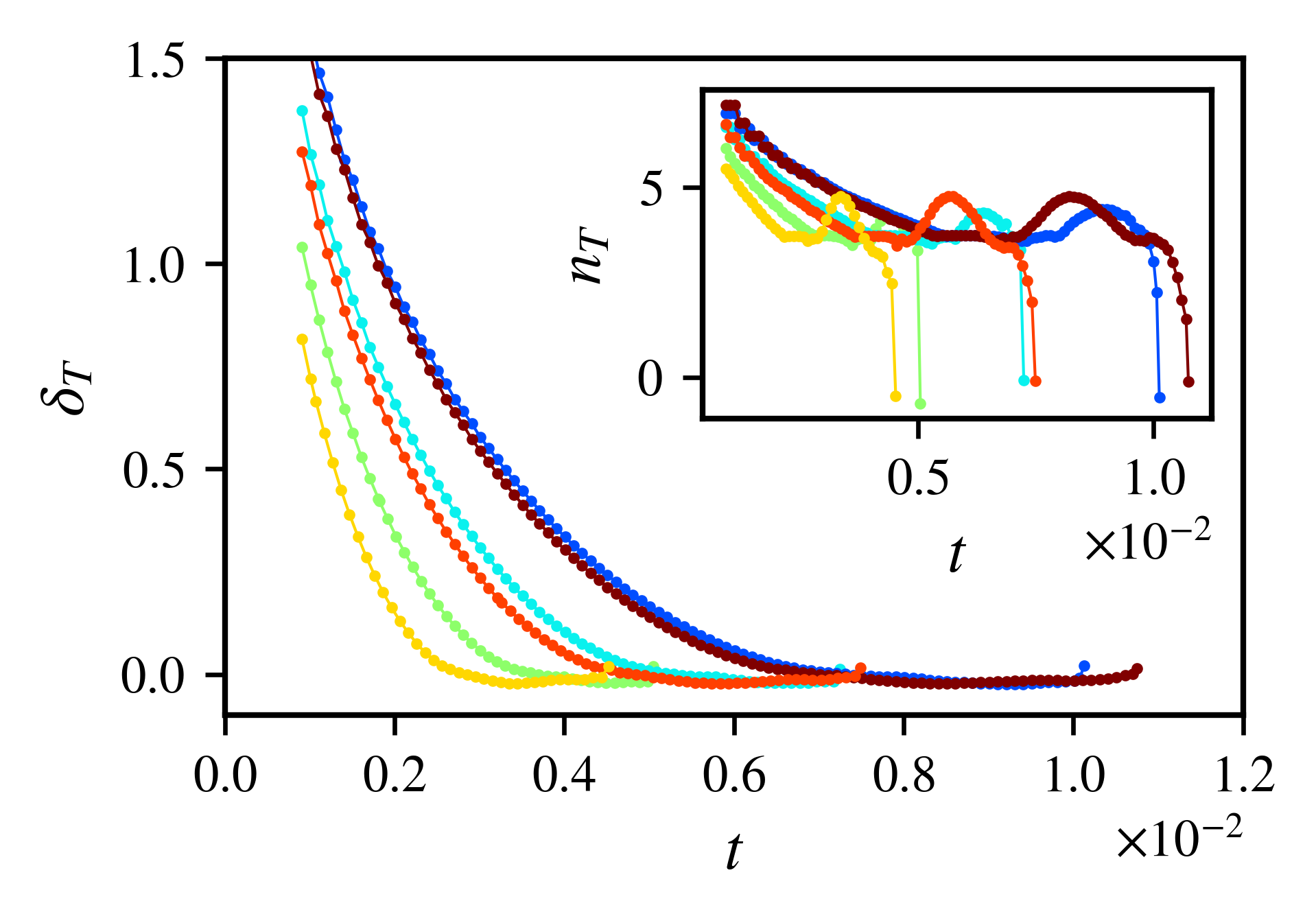}};
    \draw (-1.9,-1) node {(a)};
    \end{tikzpicture}
    \begin{tikzpicture}
    \draw (0,0) node[inner sep=0]{\includegraphics[height=0.31\linewidth]{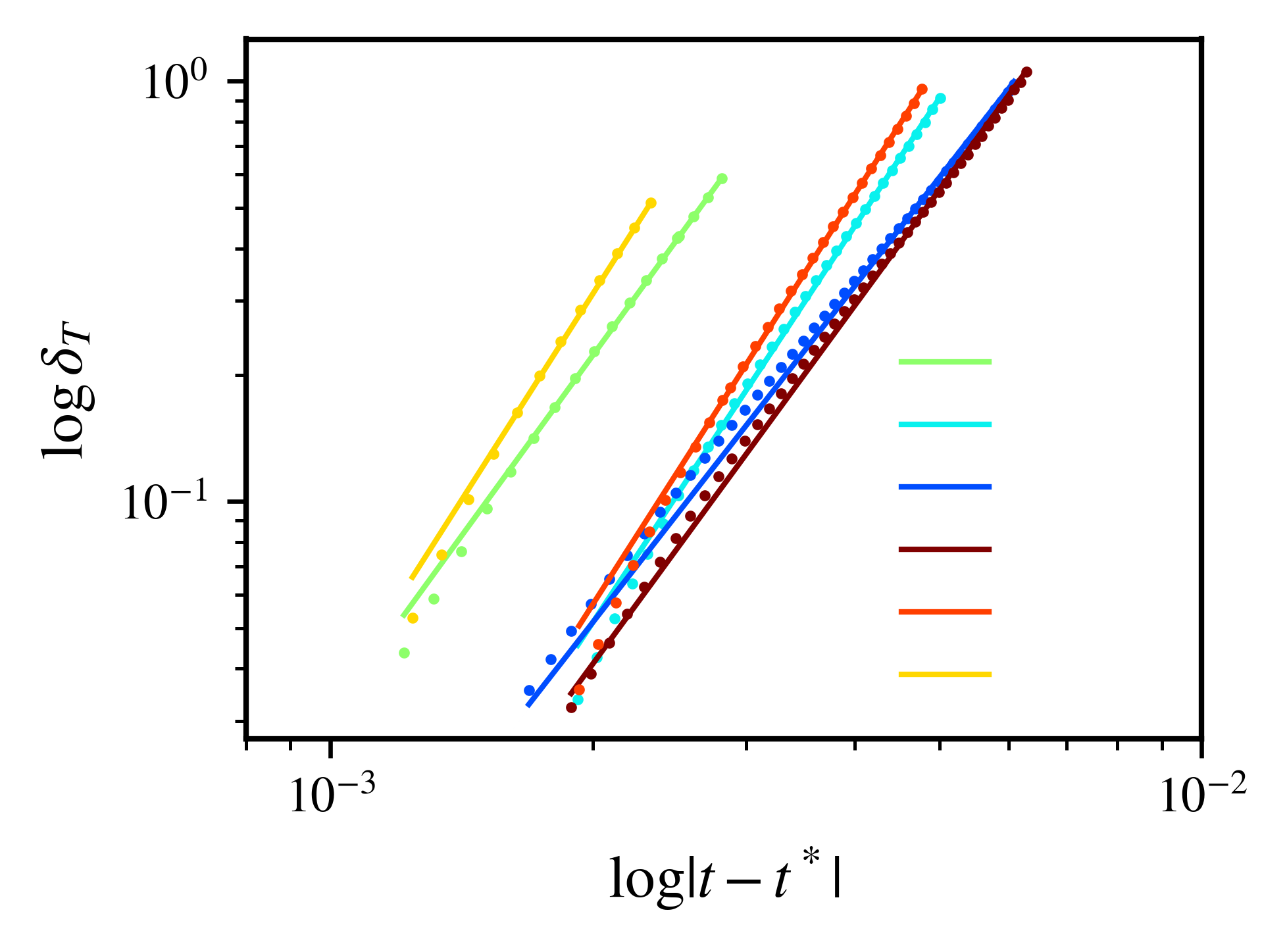}};
    \draw (-1.7,-1) node {(b)};
   \draw (2.35, 0.85-0.23) node {$\scriptstyle \mathcal{C}=0.50$};
   \draw (2.35, 0.52-0.23) node {$\scriptstyle \mathcal{C}=0.80$};
   \draw (2.35, 0.16-0.2) node {$\scriptstyle \mathcal{C}=0.90$};
   \draw (2.35,-0.19-0.19) node {$\scriptstyle \mathcal{C}=1.10$};
   \draw (2.35,-0.56-0.15) node {$\scriptstyle \mathcal{C}=1.20$};
   \draw (2.35,-0.92-0.12) node {$\scriptstyle \mathcal{C}=1.50$};
    \end{tikzpicture}
    \begin{tikzpicture}
    \draw (0,0) node[inner sep=0]{\includegraphics[height=0.3\linewidth]{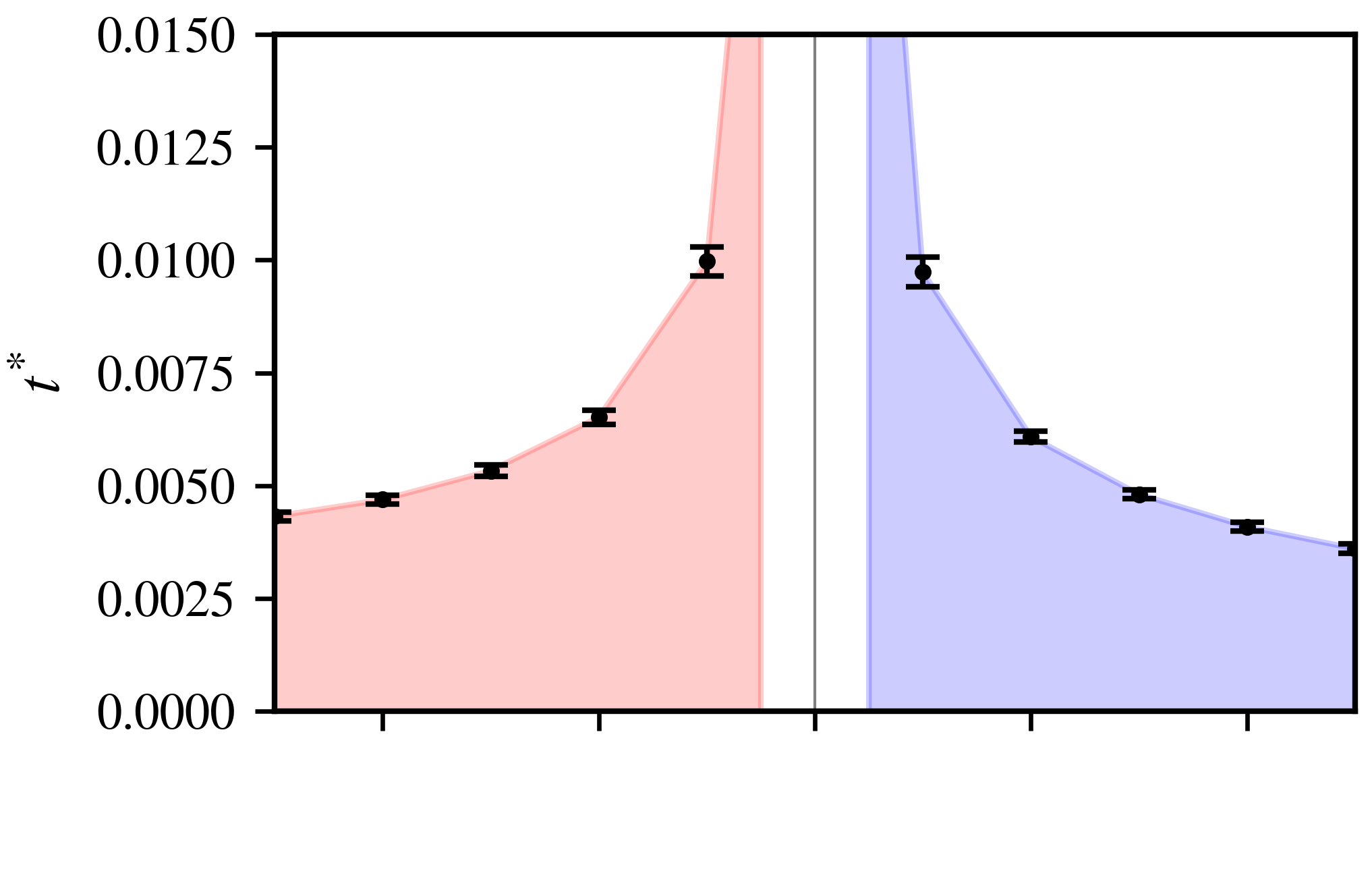}};
    \draw (-1.75,-1.1) node {(c)};
   \draw (0.75,-1.8) node { $\displaystyle 1.00 $};
   \draw (-0.5,-1.8) node { $\displaystyle 0.80 $};
   \draw (-1.68,-1.8) node { $\displaystyle 0.60 $};
   \draw (-0.5 +3.6,-1.8) node { $\displaystyle 1.40 $};
   \draw (-1.68+3.6,-1.8) node { $\displaystyle 1.20 $};
   \draw (0.75,-2.4) node {$\mathcal{C}$};
    \end{tikzpicture}
    \caption{ Plots versus $t$ of (a) the analyticity-strip width $\delta_T$ (and $n_T$ in the inset) obtained by performing the nonlinear fit~\eqref{eq:analyticity-strip} of the total energy Fourier spectrum at the wall ($r=1$) for flows initiated with different choices of $\mathcal{C}$ in the initial data. 
    (b) Log-log plot versus $|t-t^*|$ of $\log \delta_T$ and the corresponding power-law fit $\delta_T \sim |t-t^*(\mathcal{C})|^{\gamma(C)}$ for flows initiated with different choices of $\mathcal{C}$ in the initial data\noop{~\eqref{eq:ic}}. 
    (c) Plot versus $\mathcal{C}$ of the estimate for the time of singularity $t^*(\mathcal{C})$ that we obtain from the power-law fit of $\delta_T(t)$ from (a); we use $(N=256, M =512)$.}
    \label{fig:delta_c}
\end{figure}
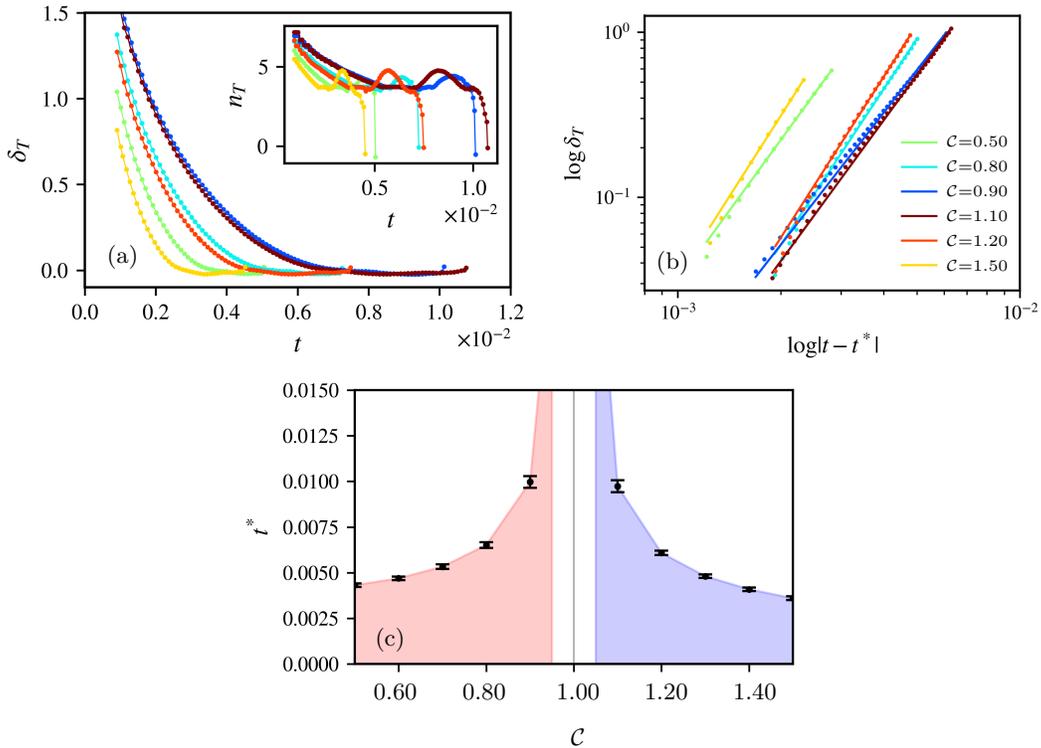

\noop{In Fig.~\ref{fig:80_spectra} (b), we plot the width of the analyticity strip $\delta_T(t)$ versus $t$ for several values of $\mathcal{C}$;
in the inset, we plot $n_T(t)$, the exponent of the power-law prefactor in Eq.~\eqref{eq:analyticity-strip}. }
Clearly \noop{both} \add{the rate of decay of} $\delta_T(t)$ \noop{and $n_T(t)$} depends on $\mathcal{C}$.
\noop{; the former governs the dominant behaviour of Eq.~\eqref{eq:analyticity-strip}, so we concentrate on it. }
In Fig.~\ref{fig:delta_c} (a), we present log-log plots of $\delta_T(t)$ versus $t$; the data from our numerical simulations are shown by filled circles;
the linear regimes in these plots are consistent with the power-law forms $\delta_T(t) \sim |t-t^*(\mathcal{C})|^{\gamma(\mathcal{C})}$, which we show via solid lines.
These plots provide strong evidence for FTSs in the \add{3D-axisymmetric} IMHD model \noop{~\eqref{eq:main}-~\eqref{eq:poisson} with the BCs~\eqref{eq:bc} and the initial data~\eqref{eq:ic}}
for \add{the flows that we initialise with} $\mathcal{C} \neq 1$. We display the $\mathcal{C}$-dependence of the singularity time $t^*(\mathcal{C})$ in Fig.~\ref{fig:delta_c} (b)~\cite{Note4}.
The case \noop{ $C=100$ (i.e., } $\mathcal{C}=1$ is special because the kinetic and magnetic terms on the RHS of Eqs.~\eqref{eq:main} cancel each other,
so the solution remains stationary and there is no singularity;
$t_*(\mathcal{C})$ decreases as $\mathcal{C}$ moves away from \add{$\mathcal{C}=1$} \noop{$C=100$}. \noop{; but the plot of $t_*(C)$ [Fig.~\ref{fig:delta_c} (b)] is not symmetrical about the vertical line $C=100$.}

\noop{Before we present our results for  $\delta_T(t)$} 
Here, we note that there are a finite number of Fourier-Chebyshev modes in any practical numerical simulation, so there is energy-backflow from high-$k$ modes, near the dealiasing cut-off $K_G$, and our Galerkin-truncated system moves towards thermalisation~\cite{di2018dynamics,tyger1}, which is reflected in the flattening of $\mathcal{S}_V(r=1,k,t)$ and $\mathcal{S}_B(r=1,k,t)$ 
in Fig.~\ref{fig:80_spectra}(b) at late times $t$. In physical space, this leads to the emergence of localised, oscillatory structures in the kinetic and magnetic fields, as we discuss below). These structures are referred to as \textit{tygers}~ \cite{kolluru2022insights,tyger1,kolluru2024novel}
; they are precursors to thermalisation~\cite{tyger1} 
and often appear just prior to an FTS~\cite{kolluru2022insights,tyger1}. 

\add{To understand the development of the singularity in physical space,} we now examine the \add{streamlines and secondary flows of the} velocity and magnetic fields \noop{in physical space}, for 
two illustrative values of \noop{$C=80$ and $C=120$ in Eq.~\eqref{eq:ic}, or, equivalently,} $\mathcal{C}=0.8$ and $\mathcal{C}=1.2$; henceforth, we refer to the resulting flows as Flow-A and Flow-B, respectively. The observations that we make for Flow-A (Flow-B), hold true for flows initialised with other choices for $\mathcal{C}$, with $\mathcal{C}<1$ ($\mathcal{C}>1$).

\begin{figure}[htbp]
    \centering
    \begin{tikzpicture}
    \draw (0,0) node[inner sep=0]{\includegraphics[width=0.45\linewidth]{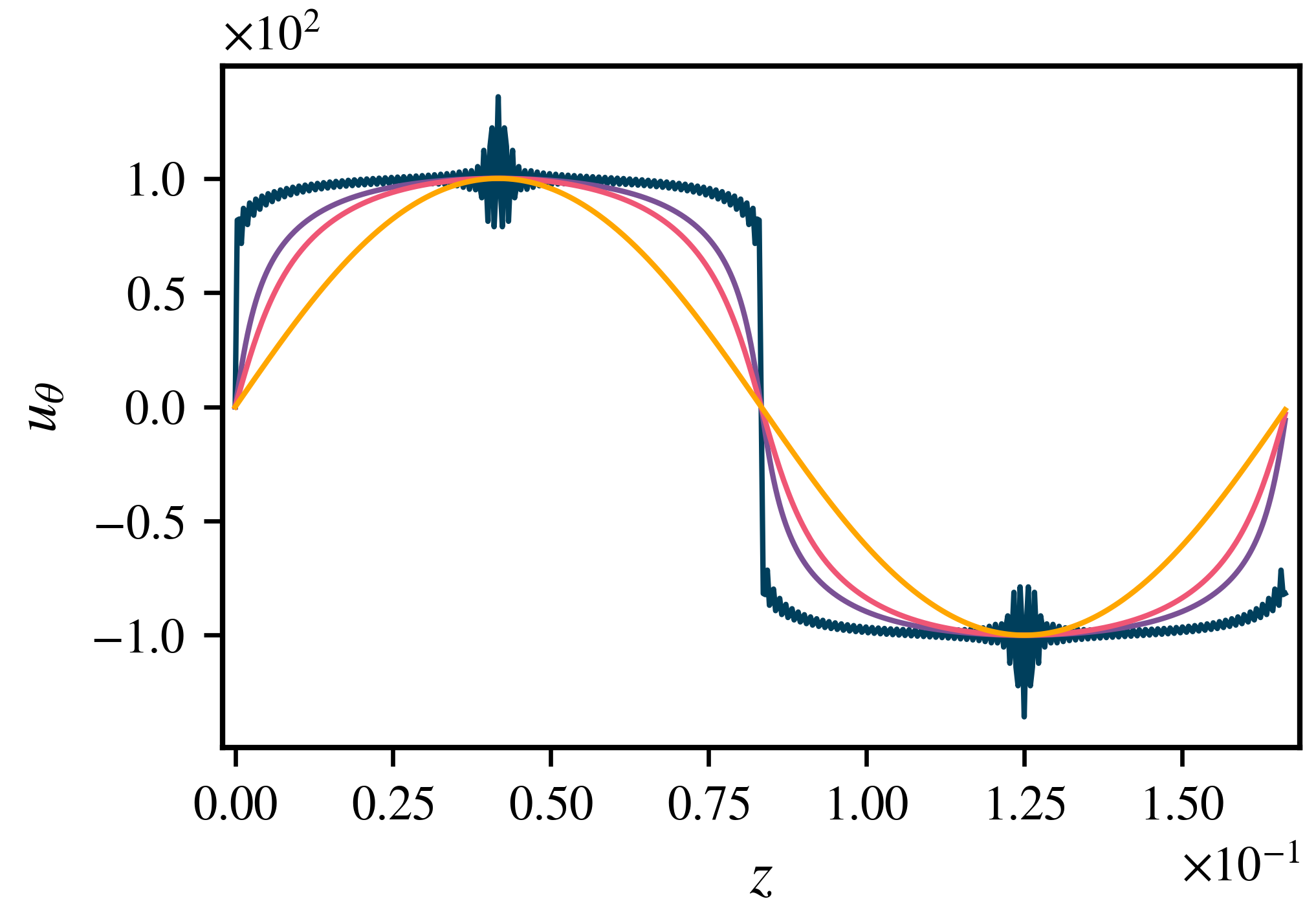}};
    \draw (-1.9,-1.2) node {(a)};
    \end{tikzpicture}
    \begin{tikzpicture}
    \draw (0,0) node[inner sep=0]{\includegraphics[width=0.45\linewidth]{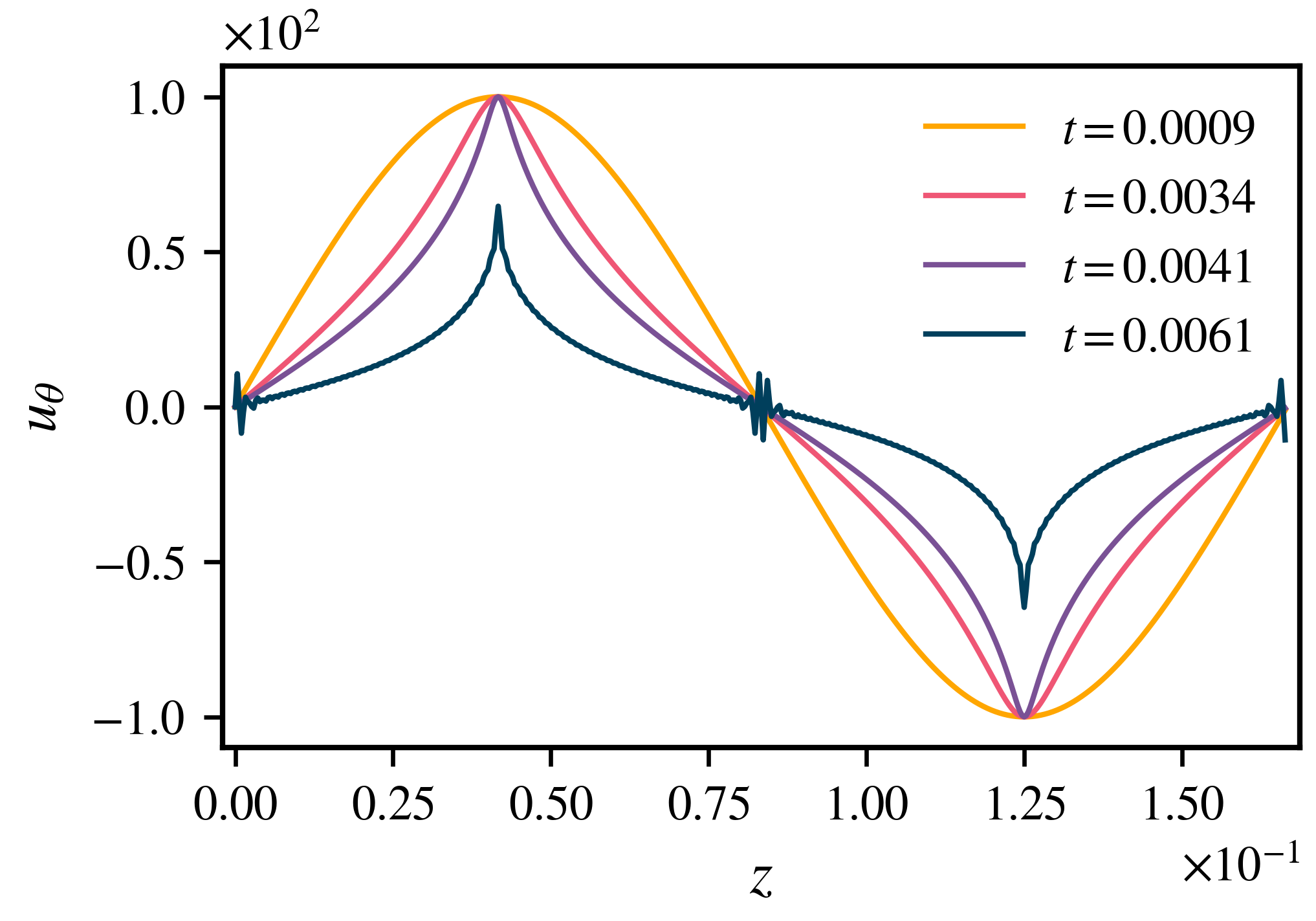}};
    \draw (-1.9,-1.2) node {(b)};
    \end{tikzpicture}    
    \caption{Plot versus $z$ of the swirl velocity $u^{\theta}(r=1,z)$ superposed for various times $t$ listed in the legend, for flows initiated with (a) $\mathcal{C}=0.80$ and (b) $\mathcal{C}=1.20$. 
    Here, we use a resolution of $(N_r=256, N_z=512)$. At later times, we see the development of localized oscillatory structures called \textit{tygers} in these fields. See the Supplementary Information for similar plots of $b^\theta$.}
    \label{fig:u_tygers}
\end{figure}

\add{\textbf{Profile at wall:}} In Flow-A, the velocity fields are initially stronger than their magnetic counterparts. 
In Fig.~\ref{fig:u_tygers}(a), \noop{we plot the swirl velocity $u^{\theta}(z)$ at the wall $(r=1)$; at early times [e.g., $t=9\cdot 10^{-4}$] } \add{we observe that at early times ($t = 9\cdot 10^{-3}$), }$u^{\theta}$ has a nearly sinusoidal profile at the wall, which develops \add{into} a square profile \add{at later times} $(t=6.1\cdot 10^{-3})$, \noop{$u^{\theta}$}. \noop{which} \add{This} in turn results in the intensification of shear at the meridional plane ($r=1,z=L/2$) and tygers appear near $z=\{ L/4, 3L/4 \}$ at this time. [For similar plots of other fields, see Fig. F2 in the Supplementary Information.] \noop{This is the first observation of tyger oscillations
in any ideal MHD system. As we have noted above, such tygers are precipitated by the loss of spectral convergence in our Galerkin-truncated scheme. Tygers were first reported in Fourier pseudospectral studies of the 1D inviscid Burgers equation in Ref.~\mbox{\cite{tyger1}} and by us in Ref.~\mbox{\cite{kolluru2022insights}} for the 3D axisymmetric Euler equations; a recent study has explored tygers in the 3D incompressible Euler equation~\mbox{\cite{murugan2020suppressing}}. These structures emerge in the smooth regions of the flow, far from the discontinuities where we would expect Gibbs oscillations; and tygers are now understood to be precursors to thermalisation in the Galerkin-truncated PDEs, which conserve the energy~\mbox{\cite{tyger1,kolluru2022insights,murugan2020suppressing}}.}%
%
%
%
In Flow-B [\noop{$C=120$, i.e., }$\mathcal{C}=1.2$] the magnetic field is initially stronger than the velocity field. In Fig.~\ref{fig:u_tygers}(b) the plots of $u^{\theta}(z)$ at the wall ($r=1$), for various representative times, show that the initial sinusoidal profile of $u^{\theta}(r=1,z)$ (yellow line) evolves into a \add{tent-like} \noop{triangular} profile (pink line) at intermediate times, losing differentiability at $z=L/4,\, 3L/4$. At later times [e.g., $t\simeq 6.1\cdot 10^{-3}$], \noop{$u^\theta$}  this is damped (blue line) as the flow decelerates along the wall, and we see the development of cusps in $u^{\theta}(r=1,z$ at $z=\{ L/4,3L/4 \}$, which are distinctly different from the square-type
profile in Fig.~\ref{fig:u_tygers}(a) for Flow-A. Tygers appear too, but away from the positions of the cusps. \noop{Such cusps, and the tygers associated with them, have not been
observed so far in the ideal hydrodynamical PDEs we have mentioned above.}

\begin{figure}[htbp]
    \centering
    \begin{tikzpicture}
    \draw (0,0) node[inner sep=0]{
    \includegraphics[width=0.9\linewidth]{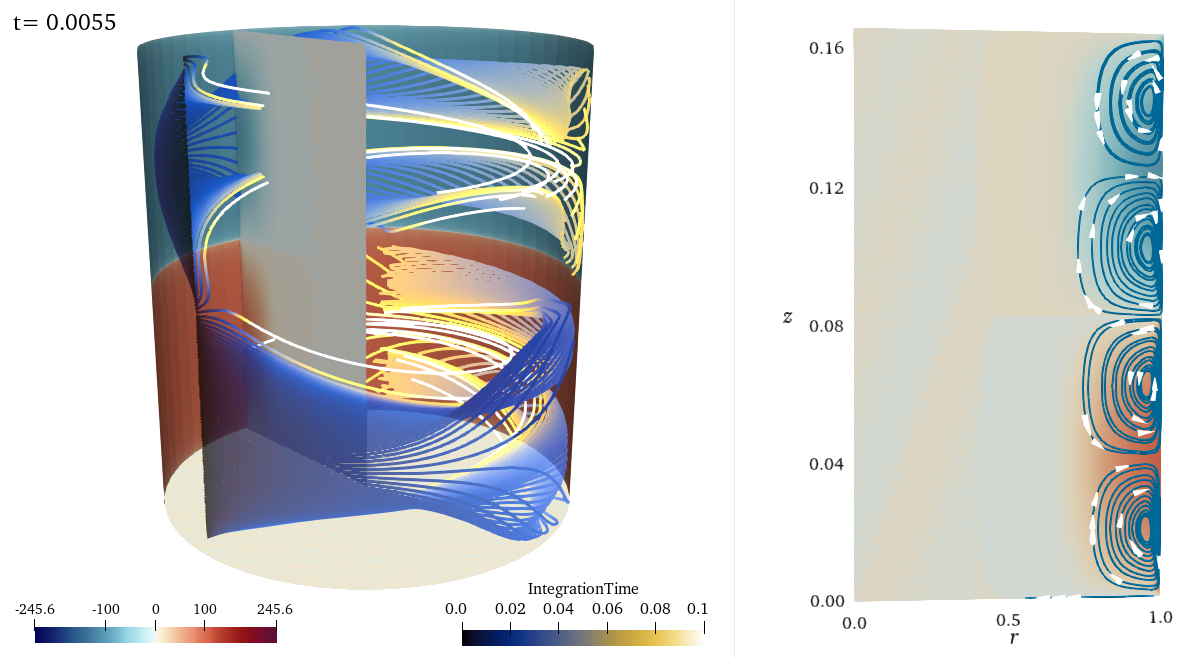}
    };
    \draw (-7,2.5) node {(a)};
    \draw (2,2.5) node {(b)};
    \draw (-5.4,-3) node {$u^{\theta}$};
    
  \def\linelength{6.725} 
  \def\segment{\linelength/4} 
  \begin{scope}[shift={(7.4,-3.2)}]
    \draw[thick] (0,0) -- (0,\linelength);
    \foreach \i/\label in {
      0/$0$, 
      1/${L/4}$, 
      2/${L/2}$, 
      3/${3L/4}$, 
      4/$L$} 
    { \pgfmathsetmacro\y{\i*\segment}
      \draw (-0.1, \y) -- (0.1, \y); 
      \node[right] at (0.15, \y) {\label};} 
  \end{scope}
  
    \end{tikzpicture}
    \caption{\textbf{Flow-A}: (a) Plot of the streamlines of the axisymmetric velocity field $\bu$ seeded from a straight line (shown on the left side, inside cylinder) at time $t=0.0055$. We plot the heat-map of $u^{\theta}(r,z)$ on the back-face of the cylinder. At early times, the streamlines (blue) are pushed outward towards the wall by the secondary flows; at later integration times, they flatten on the wall, leading to the square-type profile of $u^{\theta}(r=1,z)$ shown in Fig.~\ref{fig:u_tygers}(a).
    (b) Plots of streamlines (blue-green) of the secondary flows driven by $\bu$ in the $r-z$ plane, superposed over the heat-map of $u^{\theta}$ at the same time $t=0.0055$. We initiate the flow with $\mathcal{C}=0.80$ and use $(N=256,\, M=512)$ for this run.}
    \label{fig:80_stream}
\end{figure}
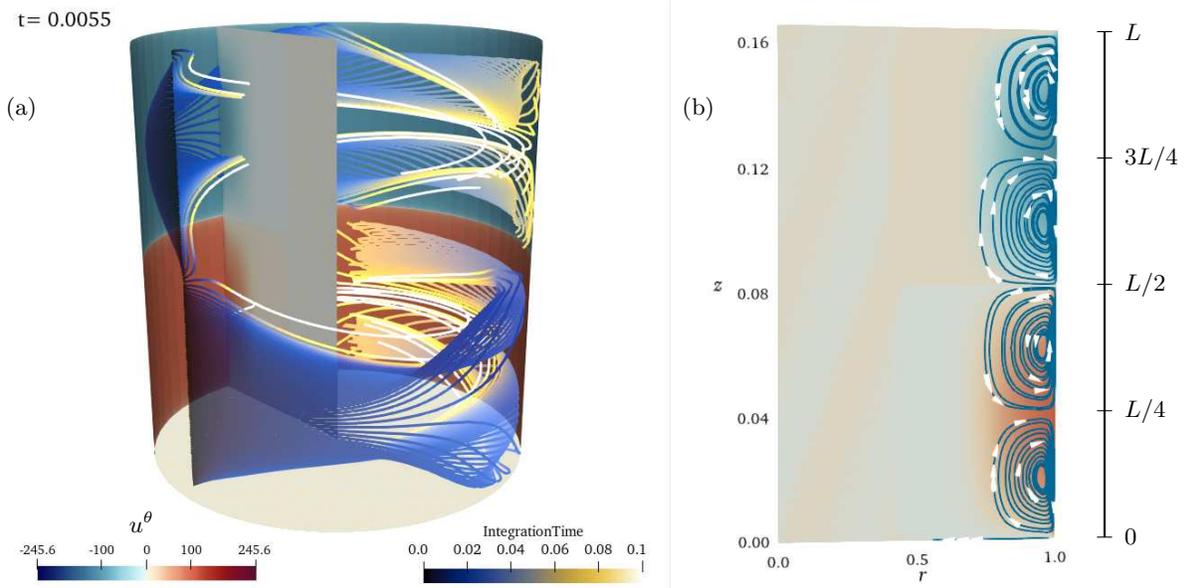

\add{\textbf{Secondary flows and streamlines:}}
\noop{The spatiotemporal evolution of $u^{\theta}$ at the wall is driven by secondary flows, which can be visualised conveniently in the $r-z$ plane.} \add{In order to understand the spatiotemporal evolution of $u^{\theta}$ at the wall described above, we now study the secondary flows in the $r-z$ plane.} We present \add{, for Flow-A,} \noop{these flows by }(i) plots of the streamlines for the \noop{axisymmetric} velocity field $\bu$, in 3D \noop{, seeded from a straight line} [Fig.~\ref{fig:80_stream} (a)] and (ii) plots of the streamlines of the secondary flow \noop{for Flow-A} in the $r-z$ plane  [blue-green solid lines in Fig.~\ref{fig:80_stream}(b)], superposed on the heat-map of $u^{\theta}$ \noop{in this $r-z$ section; both
Figs.~\ref{fig:80_stream} (a) and (b) show the fields at} \add{for} time $t=0.0055$. The secondary flow consists of the fluid rotating, within convection-type cells, along the wall; the direction of rotation of the cells is such that the fluid is driven away from $z=\{L/4,3L/4\}$ \textit{\noop{on} \add{along} the wall;}\noop{. Thus} \add{thereby}, the initial maxima of $u^{\theta}(r=1)$ \add{at these locations} \noop{at $z=\{ L/4,3L/4 \}$} are \add{similarly} advected outwards    by $u^z$.

In Fig.~\ref{fig:80_stream} (a), 
\add{we see that the streamlines, which have been started out along a straight line (blue), move into (out of) the plane given the negative (positive) sign of $u^{\theta}$ in the upper (lower) half of the cylinder.}
\noop{the streamlines from the upper (lower) half of the cylinder move into (out of) the plane given the negative (positive) sign of $u^{\theta}$ in this half. The streamlines, which have been started out along a straight line (blue),}. \add{They} bulge outwards towards the wall, because of the influence of the secondary flows, near $z=\{ L/4,3L/4\}$. 
At later times, the streamlines (yellow), which have started from near $z=\{L/4 , 3L/4\}$, fan out at the wall; this is nothing but the advection of $u^{\theta}$ outwards from $z= \{ L/4,3L/4\}$. For this time, we show the heat-map of $u^{\theta}$ on the back-face of the cylinder (colorbar of $u^{\theta}$) from which we can deduce the square-profile of $u^{\theta}(r=1,z)$ at late times [cf. Fig.~\ref{fig:u_tygers}(a)]; this leads gradually to uniform blue-green (orange) colouring of the heat-map in the upper (lower) half of the cylinder.

\add{For flows initiated with \add{$\mathcal{C}<1$} \noop{$0<C<100$}, real-space dynamics of the velocity field $\bu$}
\noop{Real-space dynamics of the velocity field $\bu$, for flows initiated with $0<C<100$,} closely follow the dynamics summarised above for Flow-A \add{with $\mathcal{C}=0.8$; however, as $\mathcal{C}$ decreases, $t_*(\mathcal{C})$ decreases.} \noop{($C=80$) [Fig.~\ref{fig:u_tygers} (a)]; of course, $t_*(C)$ decreases as $C$ moves away from $C=100$} [Fig.~\ref{fig:delta_c} (b)]. \noop{Furthermore, the potential FTS in $\bu$} \add{Importantly, the mechanism that we report here} for Flow-A is akin to that seen for the potentially singular flow in the 3D-axisymmetric wall-bounded Euler equation~\cite{luo2014potentially,kolluru2022insights,barkley} \add{($\mathcal{C}=0$ here)} \noop{; the 3DAE FTS occurs in our IMHD model when $C=0$}. Furthermore, the spatiotemporal evolution of the magnetic field $\bdb$ is similar to that of $\bu$; in particular,
we see the development of a square-profile of $b^\theta$ at the wall [see Fig. F2 in the Supplemental Information].


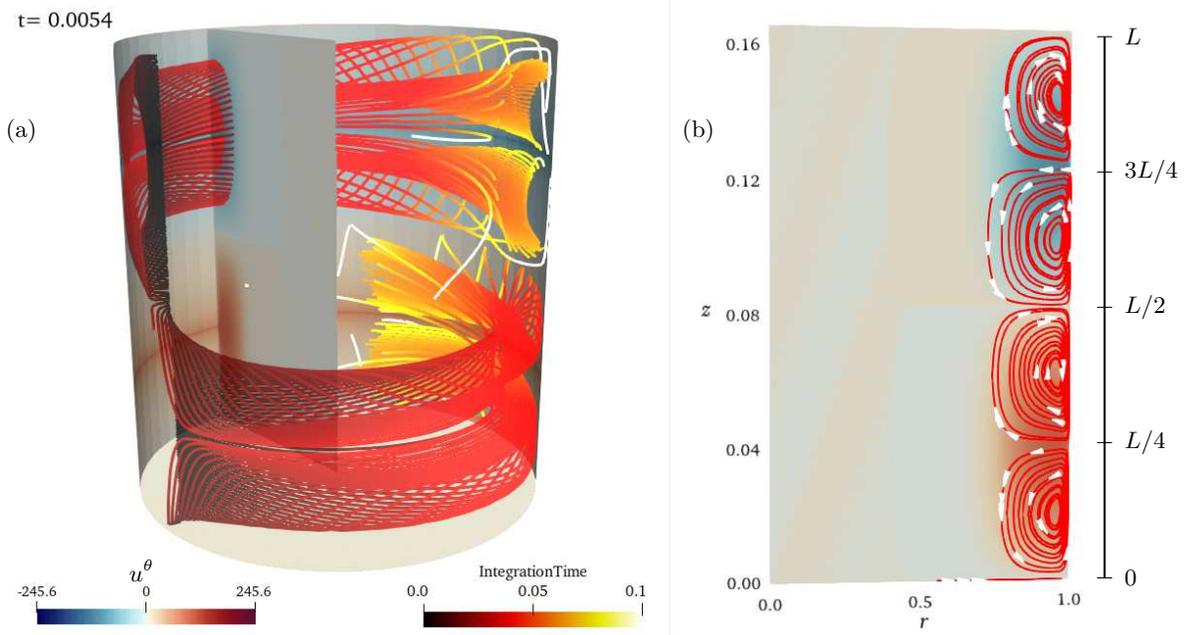
\begin{figure}[htbp]
    \centering
    \begin{tikzpicture}
    \draw (0,0) node[inner sep=0]{
    \includegraphics[width=0.9\linewidth]{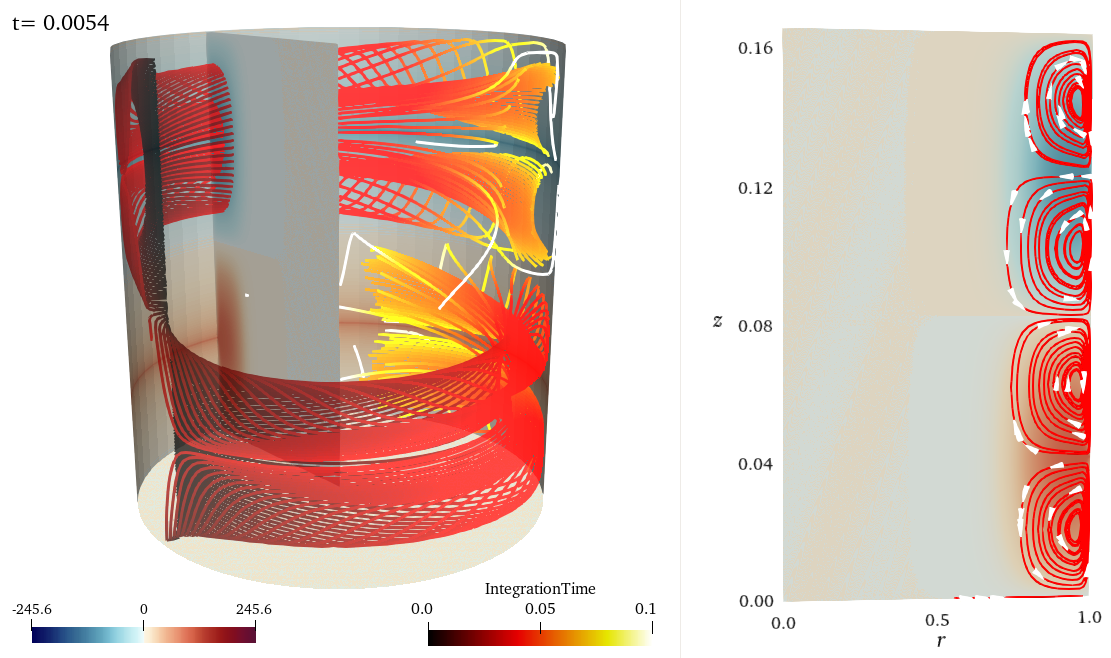}
    };
    \draw (-7,2.5) node {(a)};
    \draw (2,2.5) node {(b)};
    \draw (-5.4,-3.35) node {$u^{\theta}$};

  \def\linelength{7.2} 
  \def\segment{\linelength/4} 
  \begin{scope}[shift={(7.4,-3.45)}]
    \draw[thick] (0,0) -- (0,\linelength);
    \foreach \i/\label in {
      0/$0$, 
      1/${L/4}$, 
      2/${L/2}$, 
      3/${3L/4}$, 
      4/$L$} 
    { \pgfmathsetmacro\y{\i*\segment}
      \draw (-0.1, \y) -- (0.1, \y); 
      \node[right] at (0.15, \y) {\label};} 
  \end{scope}
  
    \end{tikzpicture}
    \caption{\textbf{Flow-B}: (a) Plot of the streamlines of the axisymmetric velocity field $\bu$ seeded from a straight line (shown on the left side, inside cylinder) at time $t=0.0054$. We plot the heat-map of $u^{\theta}(r,z)$ on the back-face of the cylinder. 
    The streamlines (red) are twisted and pulled inward by the secondary flows at $z=\{L/4,3L/4\}$, close to the singularity time $t^*(\mathcal{C})$ \add{and this leads to the cusp-like profile of $u^{\theta}(r=1,z)$ shown in Fig.~\ref{fig:u_tygers}(b)}.   (b) Plots of streamlines (red) of the secondary flows driven by $\bu$ in the $r-z$ plane, superposed over the heat-map of $u^{\theta}$ at the same time $t=0.0054$. We initiate the flow with $\mathcal{C}=1.20$ and use $(N=256,\, M=512)$ for this run.}
    \label{fig:120_stream}
\end{figure}

For Flow-B \noop{too}, we \add{now discuss} \noop{visualise} the secondary flows in the $r-z$ plane \add{that lead to the formation of the tent-like profile of $u_{\theta}(r=1,z)$ in Fig.~\ref{fig:u_tygers}. As for Flow-A, we visualise} \noop{via} (i) \noop{plots of} the streamlines for the axisymmetric velocity field $\bu$ [Fig.~\ref{fig:120_stream} (a)] and (ii) \noop{plots of} the streamlines of the secondary flow \add{in the $r-z$ plane} [Fig.~\ref{fig:120_stream} (b)], \noop{superposed on the heat-map of $u^{\theta}$ in this $r-z$ section; both
Figs.~\ref{fig:120_stream} (a) and (b) show the fields} at time $t=0.0054$.

{The red streamlines [Fig.~\ref{fig:120_stream} (b)] of the secondary flow in the $r-z$ plane, superposed on the heat-map of $u^{\theta}$, show rotating cells of fluid, similar to those for Flow-A [Fig.~\ref{fig:80_stream}(b)],
but with the opposite direction of rotation. Here, the secondary flow is such that the fluid is driven \textit{towards} $z=\{L/4,3L/4 \}$ \textit{on the wall}. This leads to the pinching of the initially sinusoidal profile of $u^\theta$, because of its advection by $u^z$, which now points inward at $z=\{L/4,3L/4\}$ in these cells.}

{The deceleration of the flow along the wall can be seen clearly in the streamline plot  [Fig.~\ref{fig:120_stream}(a)], which is seeded as in Fig.~\ref{fig:80_stream} (a). \add{Here,} Flow-B is distinctly different from
Flow-A: the sheet formed by the streamlines in each half of the cylinder, folds in on itself at $z=\{L/4,3L/4 \}$ (dark red) because of the advection of the slower-moving fluid inward, at these points, by the secondary flow. 
Thus, at later times (yellow), the streamlines seeded near $z=\{0,L/2\}$ move towards $z=L/4$ in the upper half of the cylinder; we see similar dynamics in the lower half where the fluid rotates with positive swirl velocity. Correspondingly, the heat-map of $u^\theta$, shown on the back-face of the cylinder, displays an intensification of velocity solely at $z=\{L/4,3L/4 \}$ [in contrast with the heat-map of Fig.~\ref{fig:80_stream}(a), where the flow is accelerated everywhere along the wall].  Thus, our examination of secondary flows elucidates the magnetic-field-induced suppression of the swirl velocity at the wall and the formation of cusps in Flow-B.} 

\add{\textbf{Role of pressure:}} We now demonstrate that the pressure plays a crucial role in establishing the secondary flows \add{discussed above, which in turn lead to the formation of FTS in the IMHD flows.} We
obtain the pressure from the divergence of Eq.~\eqref{eq:imhd_u} that yields the Poisson-type equation~\eqref{eq:mhd_pressure}.
For our IMHD flows, we define the effective pressure $P = p+ \tfrac{1}{2} |\bdb|^2$, which we compute for Flow-A and Flow-B using the Fourier-Chebyshev Tau Poisson solver developed by us~\cite{kolluru2022insights}.

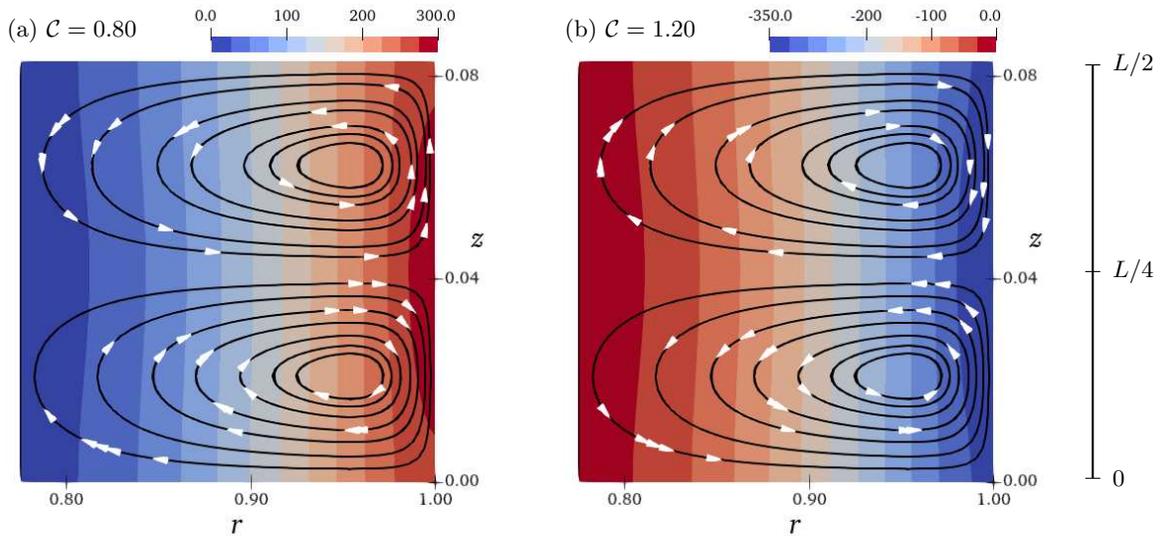
\begin{figure}[htbp]
    \centering
    \begin{tikzpicture}
    \draw (0,0) node[inner sep=0]{\includegraphics[trim={12cm 2cm 10cm 2cm},clip,width=0.43\linewidth]{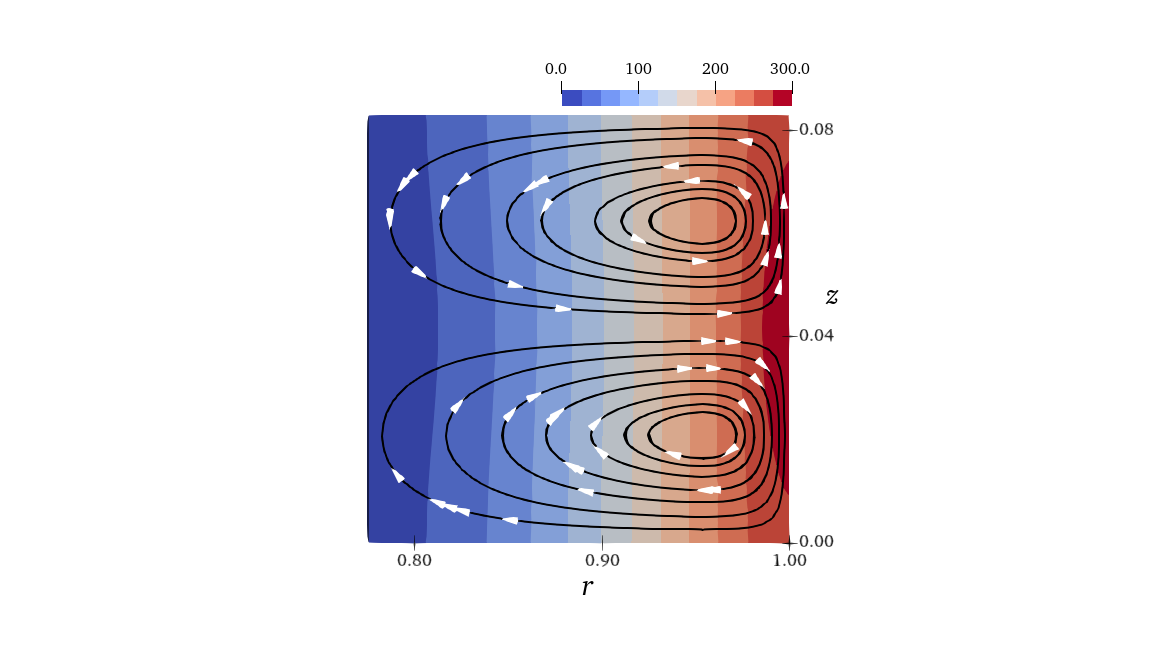}};
    \draw (-2.4,3.2) node {(a) $\mathcal{C}=0.80$};
    \end{tikzpicture}
    \hspace{1em}
    \begin{tikzpicture}
    \draw (0,0) node[inner sep=0]{\includegraphics[trim={12cm 2cm 10cm 2cm},clip,width=0.43\linewidth]{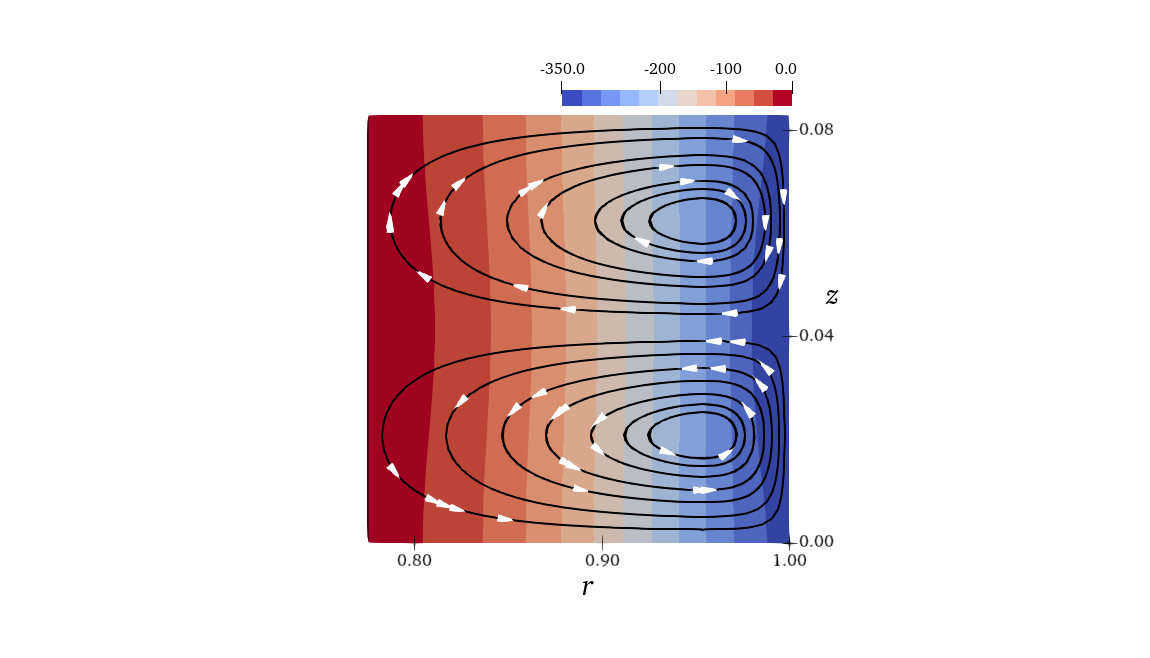}};
    \draw (-2.4,3.2) node {(b) $\mathcal{C}=1.20$};
  \def\linelength{5.5} 
  \def\segment{\linelength/2} 
  \begin{scope}[shift={(3.75,-2.75)}]
    \draw[thick] (0,0) -- (0,\linelength);
    \foreach \i/\label in {
      0/$0$, 
      1/${L/4}$, 
      2/${L/2}$} 
    {
      \pgfmathsetmacro\y{\i*\segment}
      \draw (-0.1, \y) -- (0.1, \y); 
      \node[right] at (0.15, \y) {\label}; 
    }
  \end{scope}
  
    \end{tikzpicture}
    \caption{Streamline plots of the secondary flows in the $r-z$ half-plane superposed on the heat-map of the effective pressure $P$ for (a) Flow-A $\mathcal{C}=0.80$ and (b) Flow-B $\mathcal{C}=1.20$ at time $t=1\cdot 10^{-4}$. We see that the flow moves outward from the pressure maximum at $(r=1,z=L/4)$ along the wall for Flow-A. For Flow-B, the flow moves into the pressure minimum at the same point.}
    \label{fig:secflow}
\end{figure}

In Figs.~\ref{fig:secflow} (a) and (b), we show, respectively, the heat-maps of $P$\add{$(r,z)$}, computed at the early time $t=10^{-4}$ close to the wall, in the lower half ($r\in[0.7,1]$, $z\in[0,1/12]$) of the cylinder, for (a) Flow-A and (b) Flow-B; we superpose the streamlines of the corresponding secondary flows \noop{on these heat-maps}.

\add{From }Fig.~\ref{fig:secflow} (a) \add{we can deduce} \noop{shows} that, for Flow-A, the pressure is lowest at the axis and it increases as $r\rightarrow1$; along the wall $r=1$, a local pressure maximum develops at $z=L/4$. As a result, we see that the fluid moves \textit{along} the pressure gradient on the wall, i.e., it moves outwards \noop{along the wall} from the pressure maximum at $z=L/4$ \add{in the $z$-direction}; and, in the $r$-direction, it moves away from the wall at $z=\{0,L/2\}$. At $z=L/4$, the fluid is forced to move against the local pressure gradient by virtue of incompressibility, thereby enforcing the direction of rotation of the fluid in the cells. 
Thus, the secondary flow rotates anti-clockwise for $z>L/4$ and clockwise for $z<L/4$. 
\noop{Note that the opposite sense of rotation is less energetically favoured when the fluid has to move against the local pressure gradient for a larger fraction of the path shown in Fig.~\ref{fig:80_stream} (a).}

In contrast, for Flow-B [Fig.~\ref{fig:secflow} (b)], the effective pressure $P$ is lower at the wall and increases as $r \rightarrow 0 $; now, there is a local minimum of $P$ at $z=L/4$ [cf. the maximum of $P$ for Flow-A in Fig.~\ref{fig:secflow} (a)]. As a result, the fluid moves towards $z=L/4$ along the wall, and then it moves outward radially, towards the wall at $z=\{0,L/2\}$. \noop{Here, the flow is in tandem with the local gradient of $P$. At $z=L/4$, the flow opposes this local pressure gradient and is directed inward radially, towards the axis; however, this is energetically favourable, so the resulting secondary flows rotate clockwise, for $z>L/4$, and anti-clockwise, for $z<L/4$.} 



We can relate the pressure field, which develops at initial times, to the initial data \noop{~\eqref{eq:ic}} by considering the radial-pressure balance at $r=1$ (on the wall), as shown in Eq.~\eqref{eq:pressurebalance_mhd}. 
%
\add{In order to obtain the radial force-balance, we consider the $r$-component of the 3DA-IMHD equations~\eqref{eq:main} and we assume that the flow is instantaneously stationary right before initialisation; thus, the LHS vanishes and we get, along $\hat{e}_r$: 
\begin{align}
    \underbrace{-\partial_r \left( p + \frac{(b^\theta)^2}{2} \right)}_{F_P} + & \ \underbrace{\frac{(u^\theta)^2 - (b^\theta)^2}{r}}_{F_R} = 0 \,, \qquad \qquad \text{at $r=1$ and $t=0$}.
    \label{eq:pressurebalance_mhd}
\end{align}}

We find that the gradient of $P\equiv [p + \tfrac{1}{2}(b^\theta)^2]$ has to balance the
centrifugal force [$(u^\theta)^2/r$ that points outward along $\hat{e}_r$], \textit{minus} the curvature forces 
because of the tension of the magnetic field lines [$(b^\theta)^2/r$, inward along $\hat{e}_r$], at the wall. 
For Flow-A, $\mathcal{C}=0.8$ and $b^{\theta}_0 (r,z) = 0.8 \ u^{\theta}_0(r,z)$ \noop{from Eq.~\eqref{eq:ic}}; thus, the centrifugal force is dominant, giving a resultant force $F_R$ that points outwards along $\hat{e}_r$. In order to achieve force-balance, $P$ is set up in a way such that its radial gradient points outward along $\hat{e}_r$ to balance the resultant force above [note the sign in $F_P$ in Eq.~\eqref{eq:pressurebalance_mhd}]. This is verified by the heat-map of $P$ shown in Fig.~\ref{fig:secflow} (a), with $P$ increasing as $r\rightarrow1$; the pressure maximum occurs at $z\rightarrow L/4$ as $(u^{\theta})^2-(b^{\theta})^2$ and assumes its largest value here.

{By contrast, for Flow-B where $\mathcal{C}=1.20$, the magnetic-tension forces dominate over the centripetal forces, and the resultant force $F_R$ points radially inwards. The pressure field $P$ is now such that its radial gradient points inwards. This is seen in Fig.~\ref{fig:secflow} (b), where the pressure decreases as $r\rightarrow1$ and the local minimum occurs at $z=L/4$, where the $F_R$ attains its largest value.} 




\add{In conclusion,} the global-regularity of solutions of 3D hydrodynamical PDEs, such as the 3D IMHD equations, continues to challenge physicists, fluid dynamicists, and mathematicians. Numerical studies play an important role in the quest for FTSs 
in such PDEs. We have taken the first crucial step that is required for uncovering FTSs in the 3D IMHD equations: In particular, we have presented compelling numerical evidence for the development of (potential) finite-time singularities in these equations in a wall-bounded axisymmetric domain. Furthermore, we have obtained deep insights into these FTSs by examining secondary flows and their dependence on the effective pressure $P$ \add{at initial times}.


{Our study brings out clearly the importance of emergent secondary flows, arising from differential rotation, and the role played by the effective pressure $P$ in their development.  Had the system been a viscous and resistive \textit{MHD fluid}, then the boundaries would have led to \textit{Ekman pumping}, well-studied in the contexts of magnetic stirring of liquid metals~\cite{davidson1999role,davidson2017introduction} and the persistence of geodynamos~\cite{ojha2022helicity,ranjan2024spatial}. 
In the potential FTS flows that we consider~\cite{Note5}, the differential rotation arises by virtue of the initial data.}

{The singular solutions we uncover are of two types -- either they display developing \textit{discontinuities} or 
\textit{cusps} --  depending on the relative magnitudes of the kinetic and magnetic fields at the initial time; the cusp-type singularities have not been anticipated in the types of PDEs we study.  Both these types of singularities will be of relevance and interest to plasma-physicists, fluid-dynamicists, and mathematicians. In particular, the potential FTSs we find should aid in the development of rigorous proofs for the global-regularity (or lack thereof) of solutions to the 3D IMHD PDEs, with the initial data we have proposed.}

\noop{Finally} \add{Additionally}, we \noop{present} \add{note that our work reports} the first evidence for tyger-type oscillations~\cite{tyger1,kolluru2022insights} in a Galerkin-truncated pseudospectral study of the 3D IMHD PDEs. Thus, our work has implications for the general problem of 
thermalisation in spectrally truncated ideal hydrodynamical systems. The tygers we have found are the precursors of thermalisation in the Galerkin-truncated 3D IMHD system; and they generalise considerably analogous results for ideal-fluid hydrodynamical equations~\cite{cichowlas2005effective,tyger1,di2018dynamics,kolluru2022insights,murugan2020suppressing}.

\section*{Methods}

\subsection{\add{Axisymmetric formulation of the 3D IMHD equations}}

The pressure field $p$ \add{in the IMHD equations~\eqref{eq:imhd_full}} can be eliminated using the vorticity $\bom=\nabla \times \bu$ and the current density
$\bj=\nabla \times \bdb$ and rewriting the above equations in 
the following vorticity and current formulation: 
\begin{align}
\partial_t \bom &= \nabla \times ( \bu \times \bom ) + \nabla \times ( \bj \times \bdb )\,;  \label{eq:imhd_w} \\
\partial_t \bj &= \nabla \times (\nabla \times ( \bu \times \bdb ))\,.   \label{eq:imhd_j}
\end{align}
In 3D, the stream function $\bpsi$ and magnetic potential $\bphi$ are vector quantities that are related to the primitive variables in Eqs.~\eqref{eq:imhd_u}-\eqref{eq:imhd_j} via
$\bu = \nabla \times \bpsi$ and $\bdb = \nabla \times \bphi$.
We represent vector fields in the 3D axisymmetric domain as follows: $\bu (r,z,t)=u^r(r,z,t) \ \hat{e}_r+u^{\theta}(r,z,t) \ \hat{e}_{\theta}+u^z(r,z,t) \ \hat{e}_z$, where $\hat{e}_r, \,\hat{e}_{\theta}$ and $\hat{e}_z$ are the basis vectors in the cylindrical coordinate system. We introduce the transformed variables~\cite{luo2014potentially,kolluru2022insights}
\begin{equation}
u^1 \equiv \frac{u^\theta}{r}; \; \omega^1 \equiv \frac{\omega^\theta}{r}; \; b^1 \equiv \frac{b^\theta}{r}; \; {\rm{and}}\; j^1 
\equiv\frac{j^\theta}{r}\,;
\label{eq:u1etc}
\end{equation}
and we rewrite Eq.~\eqref{eq:imhd_full} as follows:
\begin{subequations}
\begin{align}
\partial_t u^1  =& (b^r \partial_r + b^z \partial_z) b^1 - (u^r \partial_r + u^z \partial_z)u^1 -2(b^1 \partial_z \phi^1 - u^1 \partial_z \psi^1)\,; \label{eq:main1} \\
\partial_t \omega^1  =& (b^r \partial_r + b^z \partial_z) j^1 - (u^r \partial_r + u^z \partial_z)\omega^1 +  \partial_z ( (u^1)^2 - (b^1)^2)\,;\label{eq:main2} \\
\partial_t b^1  =& (b^r \partial_r + b^z \partial_z) u^1 - (u^r \partial_r + u^z \partial_z)b^1\,; \label{eq:main3} \\
\partial_t j^1  =& -(j^1 + 2\partial^2_z \phi^1)(2\partial_r u^r -\partial_z \psi^1) + (\omega^1 + 2\partial^2_z \psi^1)(2\partial_r b^r -\partial_z \phi^1) + \nonumber \\ 
&     (b^r \partial_r + b^z \partial_z) \omega^1 - (u^r \partial_r + u^z \partial_z) j^1 - b^1 \partial_z u^1 + u^1 \partial_z b^1\,. \label{eq:main4} 
\end{align}
\label{eq:main}
\end{subequations}
The Poisson equations for $\psi^1$ and $\phi^1$ are given below:
\begin{subequations}
\begin{align}
 \left( \partial_r^2 + \frac{3}{r} \partial_r + \partial_z^2 \right) {\psi^1} &= -\omega^1\,; \label{eq:pois_om_psi} \\
\left( \partial_r^2 + \frac{3}{r} \partial_r + \partial_z^2 \right) {\phi^1} &= -j^1\,. \label{eq:pois_j_phi}
\end{align}
\label{eq:poisson}    
\end{subequations}
The radial and axial components of the $\bu$ and $\bdb$ fields are obtained from $\bpsi$ and $\bphi$, respectively: 
\begin{subequations}
    \begin{align}
        u^r = -r \partial_z \psi^1\,; &\; \; \; u^z  = 2\psi^1 + r \partial_r \psi^1\,;  \\
        b^r = -r \partial_z \phi^1\,; &\; \; \; b^z  = 2\phi^1 + r \partial_r \phi^1\,. 
    \end{align}      \label{eq:main_rz} 
\end{subequations}
The solution of the above equations, in terms of the transformed variables $\{ u^1, \, \omega^1, \, ... \}$ [see Eq.~\eqref{eq:u1etc}], ensures that the corresponding fields $\{ u^{\theta}, \, \omega^{\theta}, \, ... \}$ remain smooth at the coordinate singularity on the axis of the cylinder ($r=0$).
We obtain the pressure $p$ by taking divergence of Eq.~\eqref{eq:imhd_u}, which yields the following Poisson-type equation:
\begin{subequations}
\begin{align}
    \nabla^2 \left( p + \frac{|\bdb|^2}{2} \right) &= - \nabla \cdot [(\bu \cdot \nabla) \bu] +  \nabla \cdot [(\bdb \cdot \nabla) \bdb]\,, \\
    \partial_r \left( p + \tfrac{1}{2} |\bdb|^2 \right)  &= ((u^\theta)^2 - (b^\theta)^2)\,, \qquad \qquad \text{BC at $r=1$,    $\forall (z,t)$.}
\end{align}
    \label{eq:mhd_pressure}
\end{subequations} 

\subsection{\add{Numerical method}}
We use Fourier-Chebyshev pseudospectral methods to solve Eqs.~\eqref{eq:main}. The domain $\mathcal{D}(1,L)$ is discretised using the collocation grid $\mathbf{X}_{N,M} \coloneqq \{ (r_i,z_j) : i=0,1,...,N-1 \text{ and } j=0,1,...,M-1 \}$. In the periodic $z$ direction, the collocation points $\{z_j = \tfrac{Lj}{M}: j=0,1...M-1 \}$ are distributed uniformly. In the wall-bounded $r$ direction, the collocation nodes $\{r_i = \frac{1}{2}(1+\cos(\frac{\pi(i-0.5)}{N}): i=0,1,...N-1\}$ are the roots of $T_N(2r-1)$, the order-$N$ Chebyshev polynomial of the first kind; these roots are spaced non-uniformly and they cluster near $r=0$ and $r=1$. 
Given a function $f(r,z)$, its Fourier-Chebyshev approximation on the grid $\mathbf{X}_{N,M}$ is
'
If $N=M$, the spacing between Chebyshev nodes, near $r=1$ and $r=0$, is much smaller than the spacing between the Fourier nodes.
Therefore, we use $M>N$; specifically, in our direct numerical simulation (DNS), we use $N=256$ and $M=512$, 
{which suffice to uncover the FTSs we explore and the associated tyger oscillations}. We use a fourth-order Runge-Kutta scheme, with an adaptive time step that ensures Courant–Friedrichs–Lewy (CFL) stability in both the Fourier-discretised $z$-direction and the Chebyshev-discretised $r$-direction. We compute the derivatives in spectral space and, subsequently, the nonlinear terms in physical space. In order to 
remove aliasing errors, in both Fourier and Chebyshev spectral approximations, we use the $2/3$-dealiasing method~\cite{uhlmann2000need}. We use our Tau Poisson solver~\cite{kolluru2022insights} to solve the Poisson equations \eqref{eq:pois_om_psi} and \eqref{eq:pois_j_phi} with the\noop{BCs} \add{boundary conditions.}
\noop{In Section~\ref{subsec:pressure},} We use the Tau Poisson solver for solving the pressure-Poisson problem~\eqref{eq:mhd_pressure}.


\subsection{Singularity-detection criteria}\label{subsec:Methods_detect}

We use two criteria to check for FTSs in solutions of the 3D-axisymmetric IMHD equations~\eqref{eq:main}: (a) the Caflisch-Klapper-Steele (CKS) criterion 
and (b) the analyticity-strip method~\cite{sulem1983tracing}; we discuss the details below.
\begin{itemize}
    \item[(a)] \textbf{Caflisch-Klapper-Steele (CKS) criterion:} The Beale-Kato-Majda (BKM) criterion is regularly employed to look for FTSs for the 3D Euler equations~\cite{beale1984remarks}. 
    It was generalised to ideal MHD flows by Caflisch, Klapper, and Steele~\cite{caflisch1997remarks} as follows:
    
    \textit{A solution of the 3D IMHD equations remains smooth in the time interval $[0,T_*]$ provided that}
\begin{equation}
\int^{T_*}_{0} ( ||\bom(.,t)||_{\infty} + ||\bj(.,t)||_{\infty}) \ dt < \infty.
\label{eq:cks}
\end{equation}

For our spectral simulations, we define the global maximum  $||\bom(.,t)||_{\infty} $ (resp. $||\bj(.,t)||_{\infty}$) as the maximal value of $\bom(.,t)$ (resp. $\bj(.,t)$) on the computational grid $\mathbf{X}_{N,M}$. In order to use this criterion effectively, we must capture singular structures accurately, e.g., by using
an adaptive-meshing scheme (see, e.g., Refs.~\cite{luo2014potentially,grauer1998geometry,grauer1998adaptive}).
For our fixed-grid Fourier-Chebyshev pseudospectral scheme, we employ the 
analyticity-strip method for it requires the spectral approximations of the fields; these are readily available 
to us from our DNS.

\item[(b)] \textbf{Analyticity-strip method:} This method was developed in Ref.~\cite{sulem1983tracing} for tracking complex-space singularities when Fourier pseudospectral schemes are used~\cite{sulem1983tracing,kolluru2022insights,brachet2013ideal,brachet1983small,kida1986study,brachet1992numerical,ootb,bkmas,cickptg}. At a given instant of time $t$, the real-space solution $\bu(x,t)$, of the PDE, can be analytically continued to complex-space variables $\mathfrak{z} = x + i y$, inside the analyticity-strip: $|y|<\delta (t)$. 
If at this time, the complex singularity, nearest to the real axis $\Re(x)$, is present at $\mathfrak{z}_* = x_* + i \delta$, then the Fourier spectrum of the velocity field $\hat
{u}(k,t)$ behaves as 
\begin{equation}
    |\hat{u}(k,t)| \sim |k|^{-n(t)}e^{-k\delta(t)} e^{ikx_*(t)}\,, \qquad \qquad \text{for large $k$;} \label{eq:analyticity-strip}
\end{equation}
here  $\mu$ is the exponent of the singularity and $n=\mu+1$. The presence of multiple competing singularities close to $\Re(x)$ gives rise to oscillatory behavior in the large-$k$ regime. In Ref.~\cite{kolluru2022insights}, we have extended the application of this method to pseudospectral studies in bounded domains, where we use Chebyshev discretisation in the bounded $r$ direction. Here, we employ 
our analyticity-strip method~\cite{kolluru2022insights} to analyse the Fourier-Chebyshev spectra of the velocity and magnetic fields obtained from our DNS of the 3D-axisymmetric IMHD model~\eqref{eq:main}. For our analysis, we define 
the Fourier-Chebyshev spectra $\mathcal{S}_V(r,k,t)$, $\mathcal{S}_B(r,k,t)$, and $\mathcal{S}_T(r,k,t)$, for the fluid kinetic energy, the magnetic energy, and the total energy $\mathcal{S}_T(r,k,t)$, respectively, as follows:
\begin{subequations}    
\begin{align}
\mathcal{S}_V(r,k,t) := \frac{g(k)}{2 \ N_z} \Big(& |\hat{u}^{\theta}(r,k,t)|^2 +   |\hat{u}^{r}(r,k,t)|^2 + |\hat{u}^{z}(r,k,t)|^2  \Big)\,;  \label{eq:Spectra_V} \\
\mathcal{S}_B(r,k,t) := \frac{g(k)}{2 \ N_z} \Big(& |\hat{b}^{\theta}(r,k,t)|^2 +   |\hat{b}^{r}(r,k,t)|^2 + |\hat{b}^{z}(r,k,t)|^2  \Big)\,;  \label{eq:Spectra_B} \\
\mathcal{S}_T(r,k,t) := \tfrac{1}{2} ( \mathcal{S}_V +& \mathcal{S}_B )\,;
\label{eq:Spectra} 
\end{align}
\end{subequations}
here, $g(k=0) = 1$ and $g(k>0)=2$. At a given time $t$ and radial coordinate $r$, we 
perform the nonlinear fit of the total energy spectrum $\mathcal{S}_T(r,k,t)$ using the form given in Eq.~\eqref{eq:analyticity-strip}
; we can then obtain $\delta(r,t)$ from the slope of the exponential tail in the large-$k$ regime. If at some time $t_*$, $\delta \rightarrow 0 $, then the nearest complex singularity hits the real axis, at this time, and 
the flow develops an FTS in physical space. 

Any practical implementation of a pseudospectral DNS has a finite spatial resolution, so the extraction of $\delta$ can only be accurate up until the time at which the spectral approximation is relatively free from errors~\cite{kolluru2024novel}; 
beyond this time, spectral convergence is lost, so we stop our analysis. For the 3DAE~\cite{kolluru2022insights} and for the 3D-axisymmetric IMHD flows that we discuss here, we see the emergence of localised oscillatory structures, called \textit{tygers}~\cite{tyger1}, around this time; at later times, the energy spectra show signs of thermalization~\cite{cichowlas2005effective,di2018dynamics,kolluru2022insights}.
\end{itemize}

\subsection{Conserved quantities}
\label{subsec:conserved}
The 3D IMHD equations~\eqref{eq:imhd_full} have three quadratic invariants; these are the total (kinetic + magnetic) energy $E_T$, magnetic helicity $H_M$, and the cross helicity $H_C$. For the 3D-axisymmetric system in Eq.~\eqref{eq:main}, these
are given below:
\begin{subequations}
    \begin{align}
        E_T = & \ E_V + E_B = \frac{1}{2} \int_0^1 \int_0^L |\bu|^2 + |\bdb|^2 \ r  dr dz\,; \label{eq:invariant_et}\\
        H_M =& \int_0^1 \int_0^L \bphi \cdot \bdb \ r  dr dz \,; \label{eq:invariant_hm}\\
        H_C =& \int_0^1 \int_0^L \bu \cdot \bdb \ r  dr dz\,.\label{eq:invariant_hc}
    \end{align}
\label{eq:invariants}
\end{subequations}
In Fig. F1 in the Supplementary Information, we plot versus time $t$, the numerical conservation of the invariants of Eq.~\eqref{eq:invariants} achieved using our Fourier-Chebyshev pseudospectral DNS for different values of $\mathcal{C}$ specified in the legend. In panel (a), we plot the percentage relative error in total energy $\delta E_T \%$ with respect to $t$; thus, total energy is conserved with an accuracy of $10^{-6}\%$ for early times, until $t \sim 3 \times 10^{-3}$; the accuracy is gradually lost as the system nears the potential FTS. 
In panel (b), we plot the magnetic helicity $H_M$ in Eq.~\eqref{eq:invariant_hm}, and 
in panel (c), the cross helicity $H_C$ in Eq.~\eqref{eq:invariant_hc}, with time $t$. We observe that conservation of magnetic helicity is extremely good; however, the numerical conservation of cross helicity deteriorates for $t>1\times 10^{-3}$. 

\begin{acknowledgments}
We thank T. Hertel, N. Besse, J.K. Alageshan, M. Brachet and T. Matsumoto for useful discussions; and we thank
the Anusandhan National Research Foundation (ANRF), the Science and Engineering Research Board (SERB), and the National Supercomputing Mission (NSM), India, for support,  and the Supercomputer Education and Research Centre (IISc), for computational resources.
\end{acknowledgments}

\bibliographystyle{ieeetr}
\bibliography{references}

\providecommand{\noopsort}[1]{}\providecommand{\singleletter}[1]{#1}%
\begin{thebibliography}{10}

\bibitem{doering1995applied}
C.~R. Doering and J.~D. Gibbon, {\em Applied analysis of the Navier-Stokes
  equations}.
\newblock Cambridge university press, 1995.

\bibitem{galdi2000introduction}
G.~P. Galdi, ``An introduction to the navier-stokes initial-boundary value
  problem,'' in {\em Fundamental directions in mathematical fluid mechanics},
  pp.~1--70, Springer, 2000.

\bibitem{foias2001navier}
C.~Foias, O.~Manley, R.~Rosa, and R.~Temam, {\em Navier-Stokes equations and
  turbulence}, vol.~83.
\newblock Cambridge University Press, 2001.

\bibitem{robinson2016three}
J.~C. Robinson, J.~L. Rodrigo, and W.~Sadowski, {\em The three-dimensional
  Navier--Stokes equations: Classical theory}, vol.~157.
\newblock Cambridge university press, 2016.

\bibitem{lemarie2018navier}
P.~G. Lemari{\'e}-Rieusset, {\em The Navier-Stokes problem in the 21st
  century}.
\newblock Chapman and Hall/CRC, 2018.

\bibitem{robinson2020navier}
J.~C. Robinson, ``The navier--stokes regularity problem,'' {\em Philosophical
  Transactions of the Royal Society A}, vol.~378, no.~2174, p.~20190526, 2020.

\bibitem{carlson2006millennium}
J.~A. Carlson, A.~Jaffe, and A.~Wiles, {\em The millennium prize problems}.
\newblock American Mathematical Soc., 2006.

\bibitem{Noteme}
To quote from Ref.~\cite{carlson2006millennium}: ``Show that the
  Navier–Stokes equations on Euclidean 3-space have a unique, smooth, finite
  energy solution for all time greater than or equal to zero, given smooth,
  divergence-free, initial conditions which ``decay rapidly at large
  distances.'' Alternatively, show that there is no such solution.

\bibitem{yudovich1963non}
V.~I. Yudovich, ``Non-stationary flow of an ideal incompressible liquid,'' {\em
  USSR Computational Mathematics and Mathematical Physics}, vol.~3, no.~6,
  pp.~1407--1456, 1963.

\bibitem{onsager}
L.~Onsager, ``Statistical hydrodynamics,'' {\em Nuovo Cimento}, vol.~6
  (Suppl.), p.~279–287, 1949.

\bibitem{eyink2006onsager}
G.~L. Eyink and K.~R. Sreenivasan, ``Onsager and the theory of hydrodynamic
  turbulence,'' {\em Reviews of modern physics}, vol.~78, no.~1, p.~87, 2006.

\bibitem{paicu2021global}
M.~Paicu and N.~Zhu, ``Global regularity for the 2d mhd and tropical climate
  model with horizontal dissipation,'' {\em Journal of Nonlinear Science},
  vol.~31, no.~6, p.~99, 2021.

\bibitem{biskamp2003magnetohydrodynamic}
D.~Biskamp, {\em Magnetohydrodynamic turbulence}.
\newblock Cambridge University Press, 2003.

\bibitem{choudhuri1998physics}
A.~R. Choudhuri, {\em The physics of fluids and plasmas: an introduction for
  astrophysicists}.
\newblock Cambridge University Press, 1998.

\bibitem{goedbloed2004principles}
J.~H. Goedbloed and S.~Poedts, {\em Principles of magnetohydrodynamics: with
  applications to laboratory and astrophysical plasmas}.
\newblock Cambridge university press, 2004.

\bibitem{freidberg2014ideal}
J.~P. Freidberg, {\em ideal MHD}.
\newblock Cambridge University Press, 2014.

\bibitem{galtier2016introduction}
S.~Galtier, {\em Introduction to modern magnetohydrodynamics}.
\newblock Cambridge University Press, 2016.

\bibitem{davidson2017introduction}
P.~A. Davidson, {\em Introduction to magnetohydrodynamics}, vol.~55.
\newblock Cambridge university press, 2017.

\bibitem{gurnett2017introduction}
D.~A. Gurnett and A.~Bhattacharjee, {\em Introduction to plasma physics: With
  space, laboratory and astrophysical applications}.
\newblock Cambridge University Press, 2017.

\bibitem{goedbloed2019magnetohydrodynamics}
J.~P. Goedbloed, H.~Goedbloed, R.~Keppens, and S.~Poedts, {\em
  Magnetohydrodynamics: of laboratory and astrophysical plasmas}.
\newblock Cambridge University Press, 2019.

\bibitem{brachet2013ideal}
M.~E. Brachet, M.~Bustamante, G.~Krstulovic, P.~D. Mininni, A.~Pouquet, and
  D.~Rosenberg, ``Ideal evolution of magnetohydrodynamic turbulence when
  imposing taylor-green symmetries,'' {\em Physical review E}, vol.~87, no.~1,
  p.~013110, 2013.

\bibitem{dallas2014signature}
V.~Dallas and A.~Alexakis, ``The signature of initial conditions on
  magnetohydrodynamic turbulence,'' {\em The Astrophysical Journal Letters},
  vol.~788, no.~2, p.~L36, 2014.

\bibitem{mininni2009finite}
P.~D. Mininni and A.~Pouquet, ``Finite dissipation and intermittency in
  magnetohydrodynamics,'' {\em Physical Review E—Statistical, Nonlinear, and
  Soft Matter Physics}, vol.~80, no.~2, p.~025401, 2009.

\bibitem{linkmann2015nonuniversality}
M.~F. Linkmann, A.~Berera, W.~D. McComb, and M.~E. McKay, ``Nonuniversality and
  finite dissipation in decaying magnetohydrodynamic turbulence,'' {\em
  Physical review letters}, vol.~114, no.~23, p.~235001, 2015.

\bibitem{caflisch1997remarks}
R.~E. Caflisch, I.~Klapper, and G.~Steele, ``Remarks on singularities,
  dimension and energy dissipation for ideal hydrodynamics and mhd,'' {\em
  Communications in Mathematical Physics}, vol.~184, no.~2, pp.~443--455, 1997.

\bibitem{gibbon2008euler}
J.~Gibbon, ``The three-dimensional euler equations: how much do we know?,''
  {\em Physica D: Nonlinear Phenomena}, vol.~237, no.~14-17, pp.~1894--1904,
  2008.

\bibitem{hou2009blow}
T.~Y. Hou, ``Blow-up or no blow-up? a unified computational and analytic
  approach to 3d incompressible euler and navier--stokes equations,'' {\em Acta
  Numerica}, vol.~18, pp.~277--346, 2009.

\bibitem{kerr1999evidence}
R.~M. Kerr and A.~Brandenburg, ``Evidence for a singularity in ideal
  magnetohydrodynamics: implications for fast reconnection,'' {\em Physical
  review letters}, vol.~83, no.~6, p.~1155, 1999.

\bibitem{grauer2000current}
R.~Grauer and C.~Marliani, ``Current-sheet formation in 3d ideal incompressible
  magnetohydrodynamics,'' {\em Physical Review Letters}, vol.~84, no.~21,
  p.~4850, 2000.

\bibitem{luo2014potentially}
G.~Luo and T.~Y. Hou, ``Potentially singular solutions of the {3D} axisymmetric
  {Euler} equations,'' {\em Proceedings of the National Academy of Sciences},
  vol.~111, no.~36, pp.~12968--12973, 2014.

\bibitem{barkley}
D.~Barkley, ``A fluid mechanic's analysis of the teacup singularity,'' {\em
  Proceedings of the royal society of london. Series A, Containing papers of a
  mathematical and physical character}, vol.~476, 2020.

\bibitem{kolluru2022insights}
S.~S.~V. Kolluru, P.~Sharma, and R.~Pandit, ``Insights from a pseudospectral
  study of a potentially singular solution of the three-dimensional
  axisymmetric incompressible euler equation,'' {\em Physical Review E},
  vol.~105, no.~6, p.~065107, 2022.

\bibitem{hertel2022cauchy}
T.~Hertel, N.~Besse, and U.~Frisch, ``The cauchy-lagrange method for
  3d-axisymmetric wall-bounded and potentially singular incompressible euler
  flows,'' {\em Journal of Computational Physics}, vol.~449, p.~110758, 2022.

\bibitem{Note2}
Cusp singularities have been at interfaces between different fluids, as
  discussed, e.g., in Refs.~\cite{jeong1992free,eggers2023viscous}.

\bibitem{houli}
T.~Y. Hou and R.~Li, ``Blowup or no blowup? the interplay between theory and
  numerics,'' {\em Physica D: Nonlinear Phenomena}, vol.~237, no.~14-17,
  pp.~1937--1944, 2008.

\bibitem{sulem}
C.~Sulem, P.-L. Sulem, and H.~Frisch, ``Tracing complex singularities with
  spectral methods,'' {\em Journal of Computational Physics}, vol.~50, no.~1,
  pp.~138--161, 1983.

\bibitem{Note3}
We have used the \texttt{LMFIT} package from \texttt{Python} to perform the
  nonlinear fitting.

\bibitem{Note4}
From the fit window, we exclude the range for which $\delta<\Delta z$, where
  $\Delta z$ is the smallest grid spacing in $\mathbf{X}_{256,512}$.

\bibitem{di2018dynamics}
P.~C. Di~Leoni, P.~D. Mininni, and M.~E. Brachet, ``Dynamics of partially
  thermalized solutions of the burgers equation,'' {\em Physical Review
  Fluids}, vol.~3, no.~1, p.~014603, 2018.

\bibitem{tyger1}
S.~S. Ray, U.~Frisch, S.~Nazarenko, and T.~Matsumoto, ``Resonance phenomenon
  for the galerkin-truncated burgers and euler equations,'' {\em Phys. Rev. E},
  vol.~84, p.~016301, Jul 2011.

\bibitem{kolluru2024novel}
S.~S.~V. Kolluru, N.~Besse, and R.~Pandit, ``Novel spectral methods for shock
  capturing and the removal of tygers in computational fluid dynamics,'' {\em
  Journal of Computational Physics}, vol.~519, p.~113446, 2024.

\bibitem{davidson1999role}
P.~Davidson, D.~Kinnear, R.~Lingwood, D.~Short, and X.~He, ``The role of ekman
  pumping and the dominance of swirl in confined flows driven by lorentz
  forces,'' {\em European Journal of Mechanics-B/Fluids}, vol.~18, no.~4,
  pp.~693--711, 1999.

\bibitem{ojha2022helicity}
A.~Ojha, M.~Anas, A.~Ranjan, P.~Joshi, and M.~K. Verma, ``Helicity segregation
  by ekman pumping in laminar rotating flows with gravity orthogonal to
  rotation,'' {\em Physical Review Fluids}, vol.~7, no.~3, p.~034801, 2022.

\bibitem{ranjan2024spatial}
A.~Ranjan and P.~Davidson, ``The spatial segregation of kinetic helicity in
  geodynamo simulations,'' {\em Geophysical Monograph Series}, p.~241, 2024.

\bibitem{Note5}
For the 3DAE cf. Refs.~\cite{luo2014potentially,kolluru2022insights,barkley}.

\bibitem{cichowlas2005effective}
C.~Cichowlas, P.~Bona{\"\i}ti, F.~Debbasch, and M.~Brachet, ``Effective
  dissipation and turbulence in spectrally truncated euler flows,'' {\em
  Physical review letters}, vol.~95, no.~26, p.~264502, 2005.

\bibitem{murugan2020suppressing}
S.~D. Murugan, U.~Frisch, S.~Nazarenko, N.~Besse, and S.~S. Ray, ``Suppressing
  thermalization and constructing weak solutions in truncated inviscid
  equations of hydrodynamics: Lessons from the burgers equation,'' {\em
  Physical Review Research}, vol.~2, no.~3, p.~033202, 2020.

\bibitem{uhlmann2000need}
M.~Uhlmann, {\em The need for de-aliasing in a Chebyshev pseudo-spectral
  method}.
\newblock PIK, 2000.

\bibitem{sulem1983tracing}
C.~Sulem, P.-L. Sulem, and H.~Frisch, ``Tracing complex singularities with
  spectral methods,'' {\em Journal of Computational Physics}, vol.~50, no.~1,
  pp.~138--161, 1983.

\bibitem{beale1984remarks}
J.~T. Beale, T.~Kato, and A.~Majda, ``Remarks on the breakdown of smooth
  solutions for the 3-d euler equations,'' {\em Communications in Mathematical
  Physics}, vol.~94, no.~1, pp.~61--66, 1984.

\bibitem{grauer1998geometry}
R.~Grauer and C.~Marliani, ``Geometry of singular structures in
  magnetohydrodynamic flows,'' {\em Physics of Plasmas}, vol.~5, no.~7,
  pp.~2544--2552, 1998.

\bibitem{grauer1998adaptive}
R.~Grauer, C.~Marliani, and K.~Germaschewski, ``Adaptive mesh refinement for
  singular solutions of the incompressible euler equations,'' {\em Phys. Rev.
  Lett.}, vol.~80, pp.~4177--4180, May 1998.

\bibitem{brachet1983small}
M.~E. Brachet, D.~I. Meiron, S.~A. Orszag, B.~Nickel, R.~H. Morf, and
  U.~Frisch, ``Small-scale structure of the taylor--green vortex,'' {\em
  Journal of Fluid Mechanics}, vol.~130, pp.~411--452, 1983.

\bibitem{kida1986study}
S.~Kida, ``Study of complex singularities by filtered spectral method,'' {\em
  Journal of the Physical Society of Japan}, vol.~55, no.~5, pp.~1542--1555,
  1986.

\bibitem{brachet1992numerical}
M.~Brachet, M.~Meneguzzi, A.~Vincent, H.~Politano, and P.~Sulem, ``Numerical
  evidence of smooth self-similar dynamics and possibility of subsequent
  collapse for three-dimensional ideal flows,'' {\em Physics of Fluids A: Fluid
  Dynamics}, vol.~4, no.~12, pp.~2845--2854, 1992.

\bibitem{ootb}
U.~Frisch, T.~Matsumoto, and J.~Bec, ``Singularities of euler flow? not out of
  the blue!,'' {\em Journal of statistical physics}, vol.~113, no.~5-6,
  pp.~761--781, 2003.

\bibitem{bkmas}
M.~D. Bustamante and M.~Brachet, ``Interplay between the beale-kato-majda
  theorem and the analyticity-strip method to investigate numerically the
  incompressible euler singularity problem,'' {\em Phys. Rev. E}, vol.~86,
  p.~066302, Dec 2012.

\bibitem{cickptg}
C.~Cichowlas and M.~Brachet, ``Evolution of complex singularities in kida--pelz
  and taylor--green inviscid flows,'' {\em Fluid Dynamics Research}, vol.~36,
  no.~4-6, p.~239, 2005.

\bibitem{jeong1992free}
J.-T. Jeong and H.~Moffatt, ``Free-surface cusps associated with flow at low
  reynolds number,'' {\em Journal of fluid mechanics}, vol.~241, pp.~1--22,
  1992.

\bibitem{eggers2023viscous}
J.~Eggers, ``Viscous free-surface cusps: Local solution,'' {\em Physical Review
  Fluids}, vol.~8, no.~12, p.~124001, 2023.

\end{thebibliography}


\providecommand{\noopsort}[1]{}\providecommand{\singleletter}[1]{#1}%
\begin{thebibliography}{3}%
\makeatletter
\providecommand \@ifxundefined [1]{%
 \@ifx{#1\undefined}
}%
\providecommand \@ifnum [1]{%
 \ifnum #1\expandafter \@firstoftwo
 \else \expandafter \@secondoftwo
 \fi
}%
\providecommand \@ifx [1]{%
 \ifx #1\expandafter \@firstoftwo
 \else \expandafter \@secondoftwo
 \fi
}%
\providecommand \natexlab [1]{#1}%
\providecommand \enquote  [1]{``#1''}%
\providecommand \bibnamefont  [1]{#1}%
\providecommand \bibfnamefont [1]{#1}%
\providecommand \citenamefont [1]{#1}%
\providecommand \href@noop [0]{\@secondoftwo}%
\providecommand \href [0]{\begingroup \@sanitize@url \@href}%
\providecommand \@href[1]{\@@startlink{#1}\@@href}%
\providecommand \@@href[1]{\endgroup#1\@@endlink}%
\providecommand \@sanitize@url [0]{\catcode `\\12\catcode `\$12\catcode
  `\&12\catcode `\#12\catcode `\^12\catcode `\_12\catcode `\%12\relax}%
\providecommand \@@startlink[1]{}%
\providecommand \@@endlink[0]{}%
\providecommand \url  [0]{\begingroup\@sanitize@url \@url }%
\providecommand \@url [1]{\endgroup\@href {#1}{\urlprefix }}%
\providecommand \urlprefix  [0]{URL }%
\providecommand \Eprint [0]{\href }%
\providecommand \doibase [0]{https://doi.org/}%
\providecommand \selectlanguage [0]{\@gobble}%
\providecommand \bibinfo  [0]{\@secondoftwo}%
\providecommand \bibfield  [0]{\@secondoftwo}%
\providecommand \translation [1]{[#1]}%
\providecommand \BibitemOpen [0]{}%
\providecommand \bibitemStop [0]{}%
\providecommand \bibitemNoStop [0]{.\EOS\space}%
\providecommand \EOS [0]{\spacefactor3000\relax}%
\providecommand \BibitemShut  [1]{\csname bibitem#1\endcsname}%
\let\auto@bib@innerbib\@empty
\bibitem [{\citenamefont {Berger}(1997)}]{berger1997magnetic}%
  \BibitemOpen
  \bibfield  {author} {\bibinfo {author} {\bibfnamefont {M.}~\bibnamefont
  {Berger}},\ }\href@noop {} {\bibfield  {journal} {\bibinfo  {journal}
  {Journal of Geophysical Research: Space Physics}\ }\textbf {\bibinfo {volume}
  {102}},\ \bibinfo {pages} {2637} (\bibinfo {year} {1997})}\BibitemShut
  {NoStop}%
\bibitem [{\citenamefont {Hertel}(2020)}]{hertel2020algorithme}%
  \BibitemOpen
  \bibfield  {author} {\bibinfo {author} {\bibfnamefont {T.}~\bibnamefont
  {Hertel}},\ }\emph {\bibinfo {title} {L’algorithme de Cauchy-Lagrange pour
  les {\'e}coulements axisym{\'e}triques et incompressibles d’Euler dans un
  domaine cylindrique born{\'e}: Exploration num{\'e}rique de la perte de
  r{\'e}gularit{\'e} des solutions des {\'e}quations de l’hydrodynamique par
  des m{\'e}thodes semi-Lagrangiennes d’ordres tr{\`e}s {\'e}lev{\'e}s}},\
  \href {http://www.theses.fr/2020COAZ4078/document} {Ph.D. thesis},\ \bibinfo
  {school} {Universit{\'e} C{\^o}te d'Azur} (\bibinfo {year}
  {2020})\BibitemShut {NoStop}%
\bibitem [{\citenamefont {Girault}\ and\ \citenamefont
  {Raviart}(1979)}]{girault1979finite}%
  \BibitemOpen
  \bibfield  {author} {\bibinfo {author} {\bibfnamefont {V.}~\bibnamefont
  {Girault}}\ and\ \bibinfo {author} {\bibfnamefont {P.-A.}\ \bibnamefont
  {Raviart}},\ }\href@noop {} {\emph {\bibinfo {title} {Finite element
  approximation of the Navier-Stokes equations}}},\ Vol.\ \bibinfo {volume}
  {749}\ (\bibinfo  {publisher} {Springer Berlin},\ \bibinfo {year}
  {1979})\BibitemShut {NoStop}%
\end{thebibliography}%

\end{document}


\title{Supplementary Information for \\Potential finite-time singularities of the 3D-axisymmetric ideal, incompressible, magnetohydrodynamical equations}
\author{Sai Swetha Venkata Kolluru}
 \email{saik@iisc.ac.in}
\author{Rahul Pandit}
 \email{rahul@iisc.ac.in}
\affiliation{Centre for Condensed Matter Theory, Department of Physics, Indian Institute of Science, Bengaluru, India - 560012}
\date{\today}
\maketitle
\renewcommand{\theequation}{S\arabic{equation}}
\renewcommand{\thefigure}{F\arabic{figure}}
\setcounter{equation}{0}
This Supplemental Information contains information about (a) the numerical conservation of the invariants of the 3D axisymmetric Ideal Magnetohydrodynamics (IMHD) equation [(6) in the main paper], (b) computation of magnetic helicity and (c) extra figures that show the evolution of $u_z$, $b_{\theta}$ and $b_z$ at the wall at $r=1$.

\section{Conservation of invariants}
\begin{figure}[h]
    \centering
    \begin{tikzpicture}
    \draw (0,0) node[inner sep=0]{\includegraphics[width=0.45\linewidth]{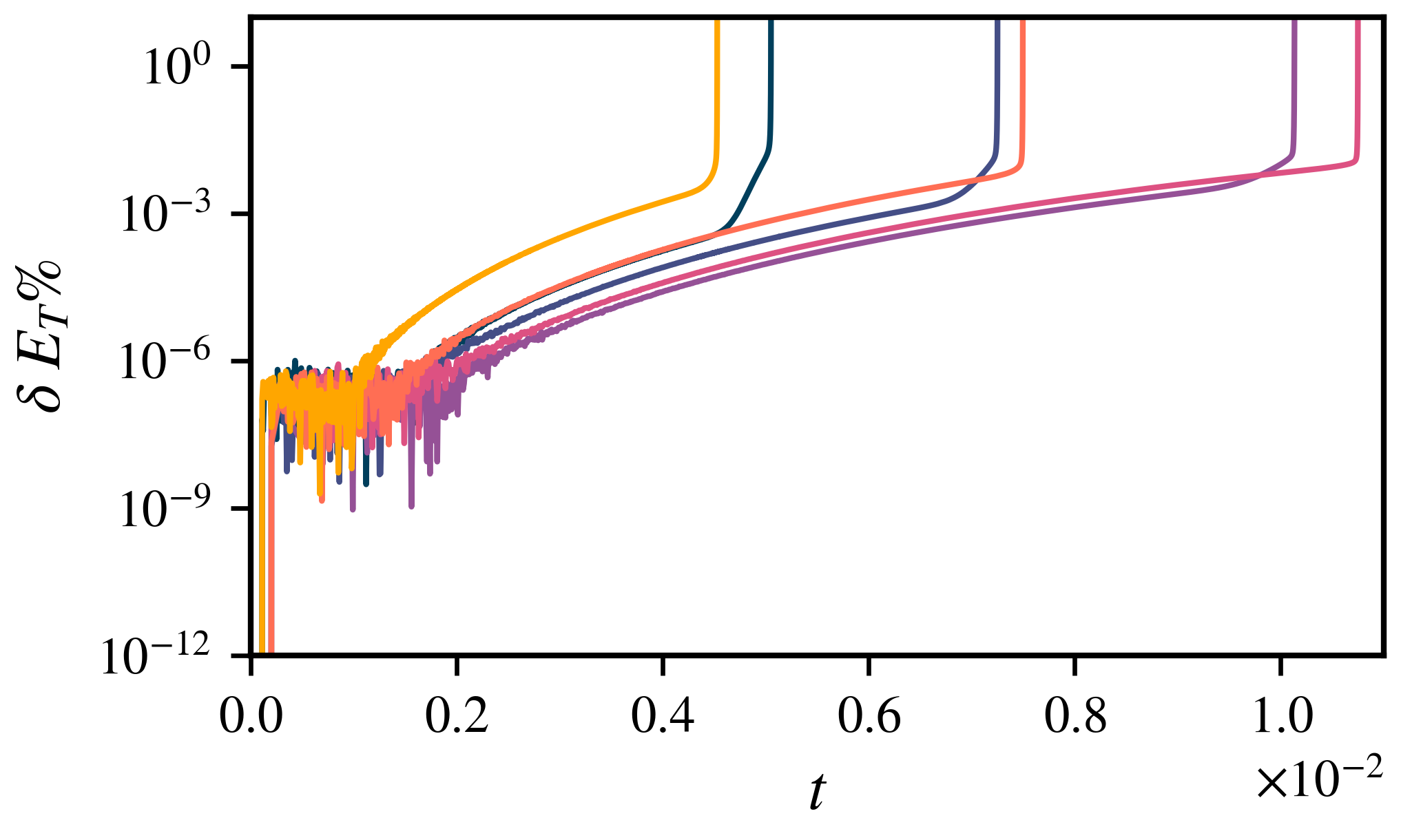}};
    \draw (3.1,-0.9) node {(a)};
    \end{tikzpicture}
    \hspace{0.03\linewidth}
    \begin{tikzpicture}
    \draw (0,0) node[inner sep=0]{\includegraphics[width=0.45\linewidth]{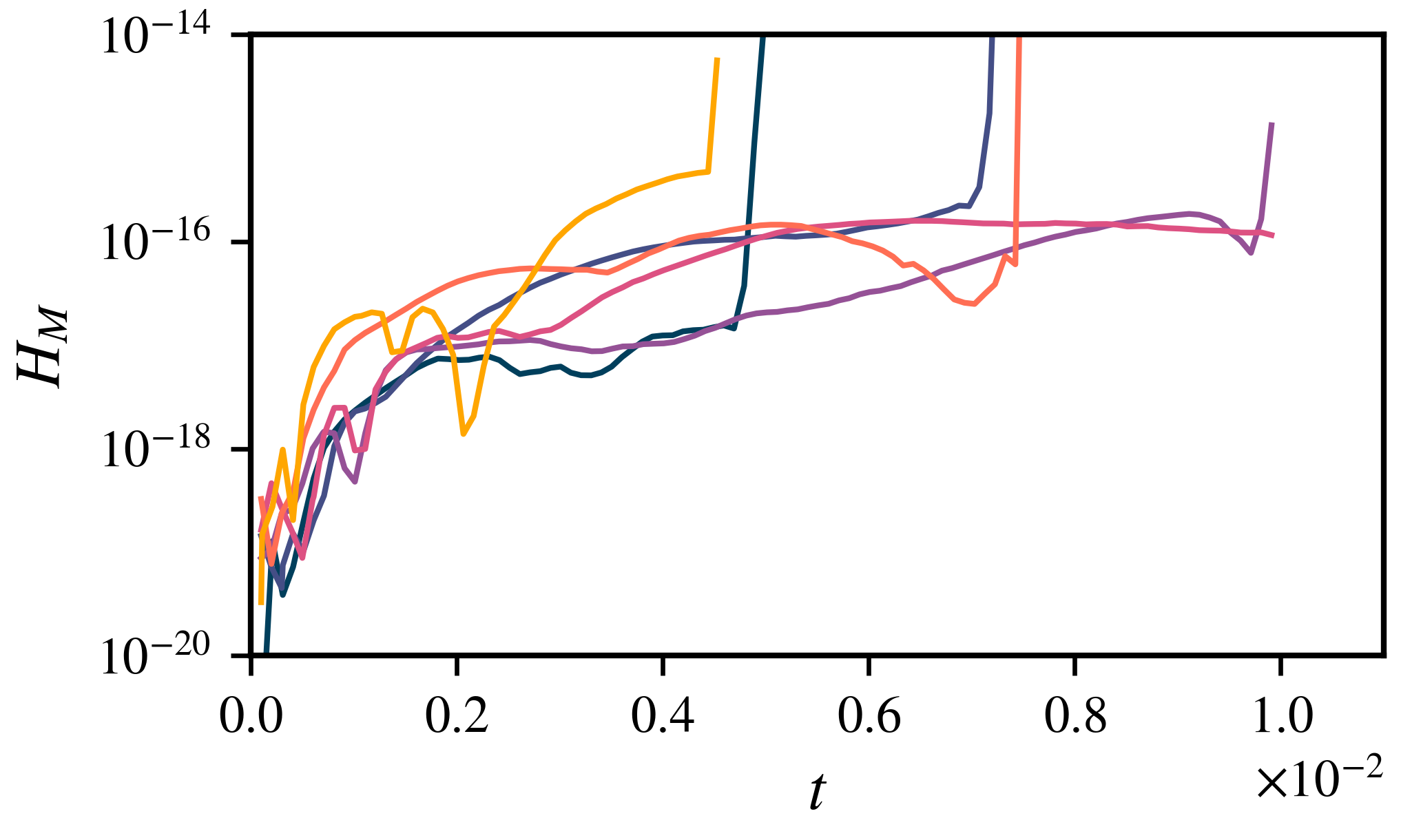}};
    \draw (3.1,-0.9) node {(b)};
    \end{tikzpicture}
    \begin{tikzpicture}
    \draw (0,0) node[inner sep=0]{\includegraphics[width=0.45\linewidth]{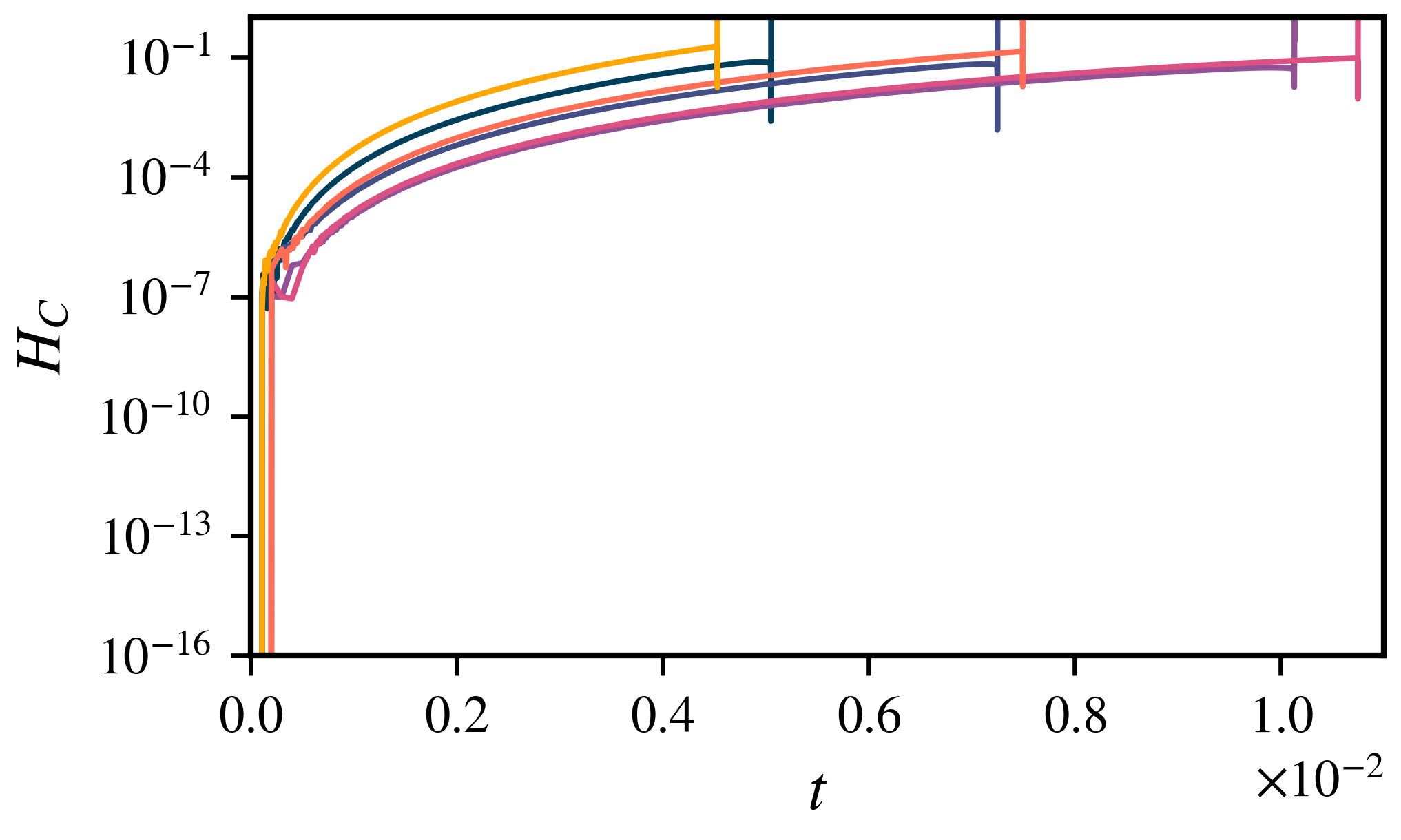}};
    \draw (3.1,-0.9) node {(c)};
    \end{tikzpicture}
    \hspace{0.04\linewidth}
    \begin{tikzpicture}
    \draw (0,0) node[inner sep=0]{\includegraphics[height=0.28 \linewidth]{figures/fig_bkm_time_v1.png}};
    \draw (-2.5,1.6) node {(d)};
    \draw (2.75,0.85) node {$\scriptstyle \mathcal{C}=0.50$};
    \draw (2.75,0.52) node {$\scriptstyle \mathcal{C}=0.80$};
    \draw (2.75,0.16) node {$\scriptstyle \mathcal{C}=0.90$};
    \draw (2.75,-0.19) node {$\scriptstyle \mathcal{C}=1.10$};
    \draw (2.75,-0.56) node {$\scriptstyle \mathcal{C}=1.20$};
    \draw (2.75,-0.92) node {$\scriptstyle \mathcal{C}=1.50$};
    \end{tikzpicture}
    \hspace{1em}
    \caption{ Plots versus $t$ of (a) the percentage change in the total energy $\delta E_T \% $, (b) the magnetic helicity $H_M$, 
    (c) the cross helicity $H_C$, and (d) the CKS integrand $\log_{10}(\log_{10}(||\bj||_{\infty}+||\bom||_{\infty}))$, for flows initialised using different choices of $\mathcal{C}$ in the initial data; these values of $\mathcal{C}$ are specified in the legend.}
   \label{fig:f1_conserved}
\end{figure}

\section{Computation of magnetic helicity}
\label{app:magheli}
Magnetic helicity is given by:
\begin{equation}
    H_M = \int_z \int_r \bphi \cdot \bdb \ r dr dz\,,
\end{equation}
where $\bphi$ is the vector-valued magnetic potential defined as $\bdb = \nabla \times \bphi$. 

In order to determine $\bphi$, we first write the components of $\bdb$ in terms of those of $\bphi$ as follows:
\begin{subequations}    
\begin{align}
    b^{\theta} &= \partial_z \phi^r - \partial_r \phi^z\,, \label{eq:btheta} \\
    b^r &= -r \partial_z \phi^1\,, \\
    b^z  &= 2\phi^1 + r \partial_r \phi^1\,; 
\end{align}
\end{subequations}
Note that we compute all components of $\bdb$ while solving Eq. (6) and (8b). From Eq. (8b) and associated boundary conditions ($b^r (r=1,z,t)=0$ and $\partial_r \phi (r=0,z,t) =0$), we solve for $\phi_\theta = r \phi^1$ using the Tau Poisson solver. 

We can determine $\phi_r$ and $\phi_z$ from $b^{\theta}$ in Eq.~\eqref{eq:btheta}. Additionally, we use the Coulomb Gauge $\nabla \cdot \bphi =0$, since the cylinder wall is considered to be a magnetic surface and since there is no mean-field in the periodic $z$-direction~\cite{berger1997magnetic}:
\begin{subequations}    
\begin{align}
\partial_z \phi^r - \partial_r \phi^z &= b^{\theta}\,, \\
\tfrac{1}{r} \phi_r + \partial_r \phi_r + \partial_z \phi_z &= 0\,.
\end{align}
\end{subequations}
We can reduce the above system of equations to a Poisson equation for $\phi_r$ given below:
\begin{subequations}
    \begin{align*}
        \left( \partial^2_r \phi_r + \partial_r \left(\frac{\phi_r}{r}\right) + \partial^2_z \phi_r  \right) = \partial_z b^\theta\,.
    \end{align*}
We can re-cast this equation in terms of $\phi_2 = \phi_r/r$ as follows:
    \begin{align}
        \left( \partial^2_r  + \frac{3}{r} \partial_r  + \partial^2_z \right) \phi_2 = \partial_z b^1\,.
    \end{align}
The above equation when supplemented with appropriate boundary conditions is a well-posed Poisson type problem which can be solved using a Tau Poisson solver akin to the one used for the solution of Eq.(8b). Once we have computed $\phi_r$, we use Eq.(6c) to compute $\phi_z$ as follows:
\begin{align}
    \partial_r \phi_z =  \partial_z \phi_r - b^\theta\,.
\end{align}
\end{subequations}
This equation can be inverted in Fourier-Chebyshev space; we require one boundary condition for this problem to be well-posed. 

\subsection*{Boundary conditions}
We now obtain boundary conditions for $\phi_r$ and $\phi_z$ at $r=0$ and $r=1$. Again, we use the Coulomb gauge $\nabla \cdot \bphi =0$:
\begin{subequations}  
\begin{alignat}{3}
   & \nabla \cdot \bphi &= 0\,, \nonumber  \\
   \implies & \tfrac{1}{r} \phi_r + \partial_r \phi_r + \partial_z \phi_z &= 0\,, \label{eq:CG} \\
    \implies & \phi_r + r(\partial_r \phi_r + \partial_z \phi_z) \; &= 0\,, \nonumber \\
    \implies & \phi_r (r=0,z,t) &= 0\,. \label{eq:bc_pr0}
\end{alignat}
\end{subequations}
For vector potential $\bphi$, we have from Refs.~\cite{hertel2020algorithme,girault1979finite}:
\begin{subequations}
\begin{align}
    \phi \times \hat{e}_r &= 0\,, \quad \text{   on }\partial D (r=1,L) \nonumber \\
    \implies \phi_z(r=1,z,t) &= 0\,. \label{eq:bc_pz1}
\end{align}
Once again using the Coulomb gauge~\eqref{eq:CG} and Eq.~\eqref{eq:bc_pz1}, we get:
\begin{align}
[ \phi_r + \partial_r \phi_r ](r=1,z,t) &= 0\,. \label{eq:bc_pr1}
\end{align}
\end{subequations}

Thus, we solve the following problems for $\phi_r$ and $\phi_z$. This allows us to compute the magnetic helicity in Eq. (13b), shown in panel (b) of Fig.~\ref{fig:f1_conserved}.

\textbf{A:} \textbf{For $\phi_r$}:
\begin{subequations} 
\begin{align*}
        \left( \partial^2_r  + \frac{3}{r} \partial_r   + \partial^2_z \right) & \phi_2 = \partial_z b^1\,, \\
    \phi_2 (r=0,z,t) = 0\,, \quad  \quad
    [2 \phi_2 +& \partial_r \phi_2](r=1,z,t) = 0\,.
\end{align*}

\textbf{B:} \textbf{For $\phi_z$}:
\begin{align*}
     \partial_r \phi_z =  \partial_z \phi_r &- b^\theta \, , \\
    \phi_z(r=1,z,t) &= 0.
\end{align*}
\end{subequations}

\newpage
\section{Extra figures}
\label{app:extras}

\begin{figure}[htbp]
    \centering
    \begin{tikzpicture}
    \draw (0,0) node[inner sep=0]{\includegraphics[width=0.45\linewidth]{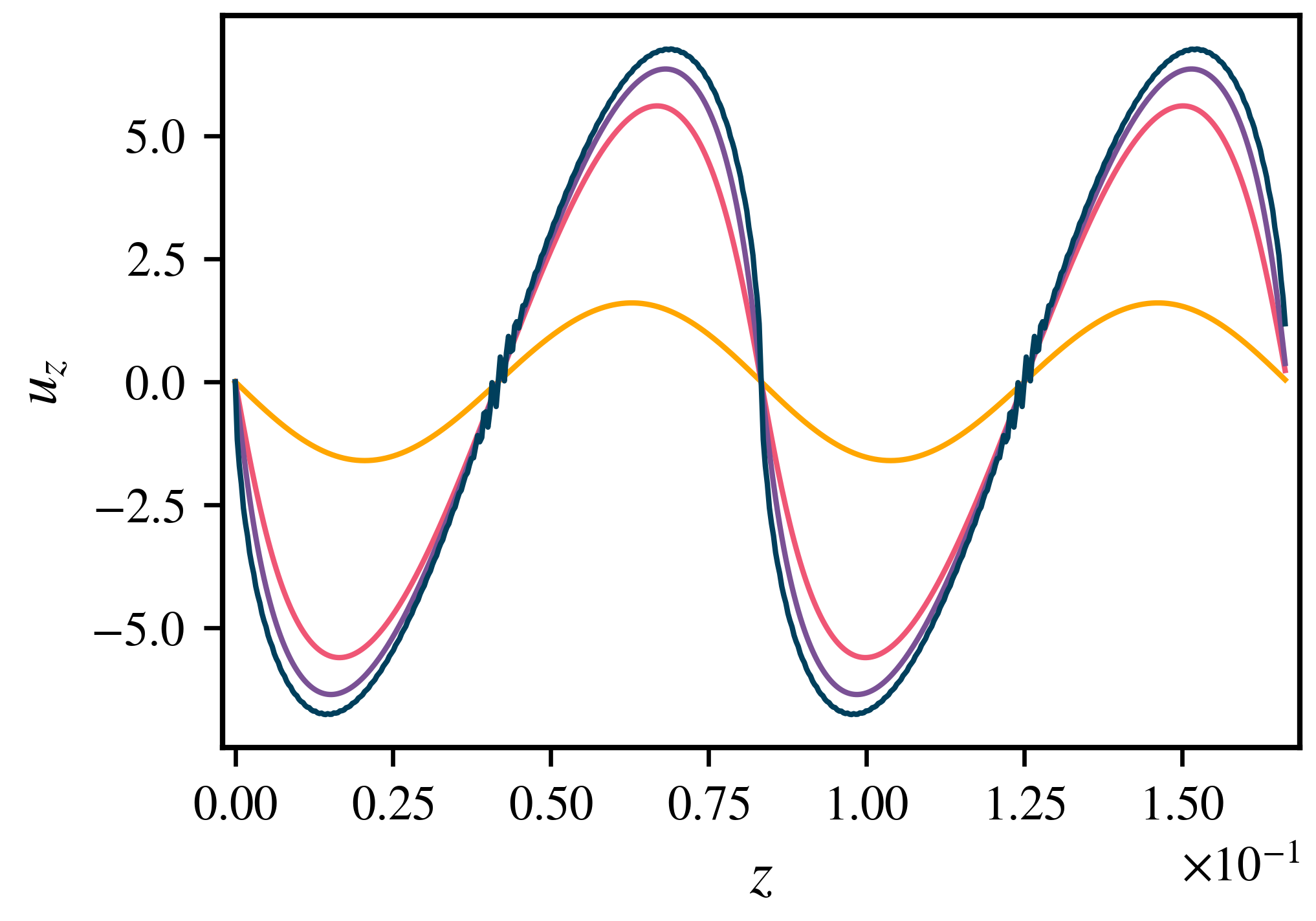}};
    \draw (-1.9,-1.2) node {(a)};
    \end{tikzpicture}
    \begin{tikzpicture}
    \draw (0,0) node[inner sep=0]{\includegraphics[width=0.45\linewidth]{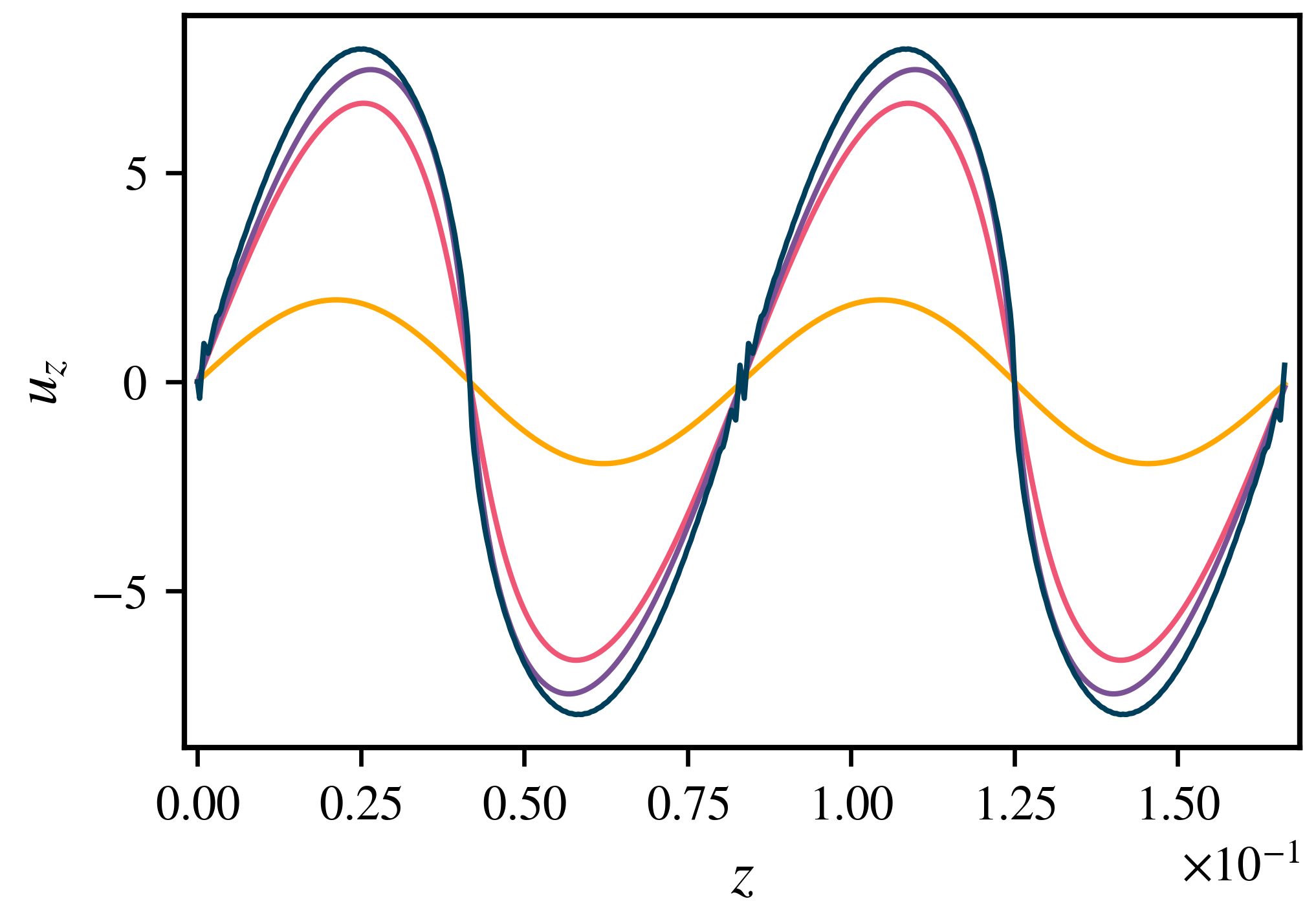}};
    \draw (-1.9,-1.2) node {(b)};
    \end{tikzpicture}    
    \begin{tikzpicture}
    \draw (0,0) node[inner sep=0]{\includegraphics[width=0.45\linewidth]{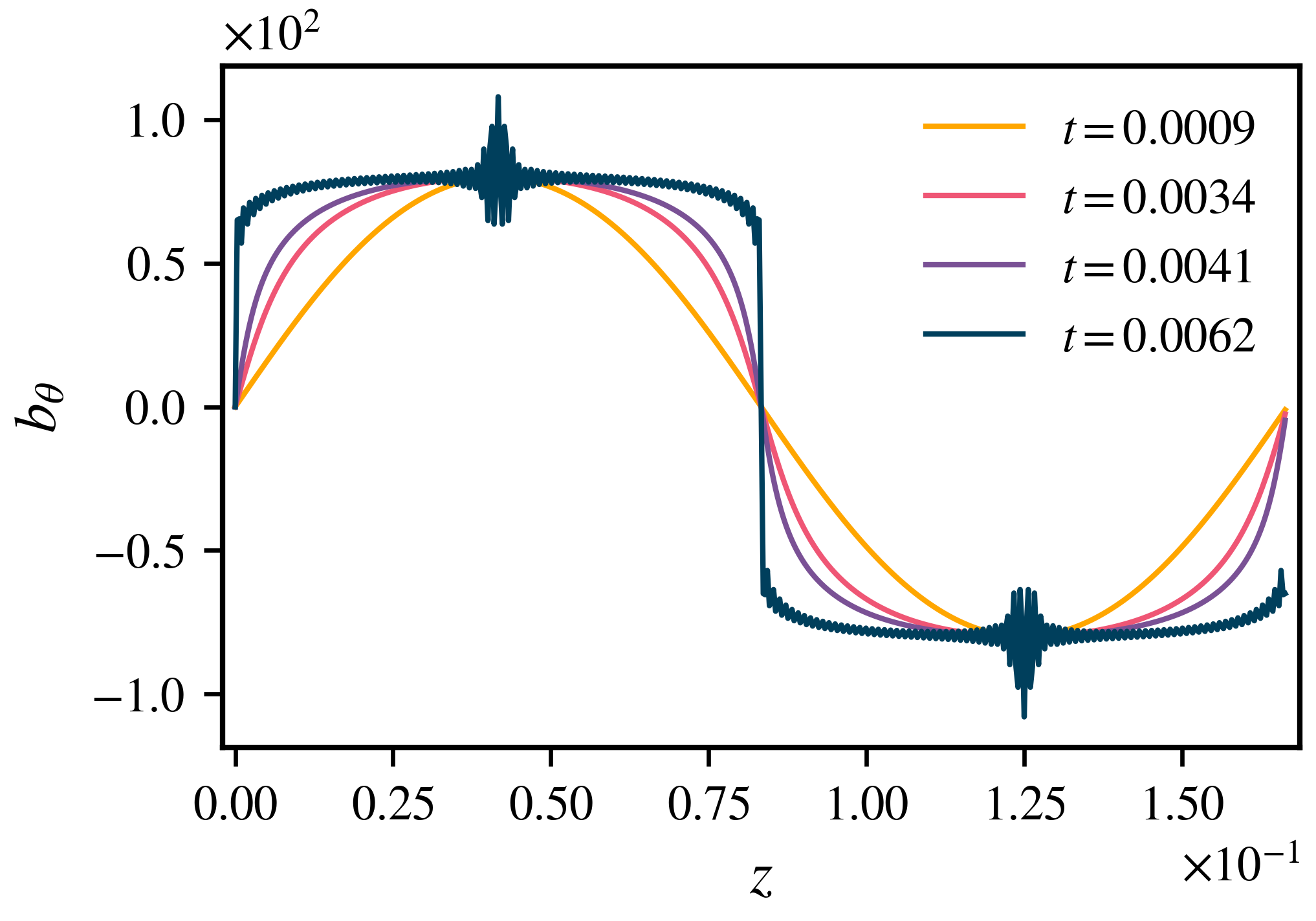}};
    \draw (-1.9,-1.2) node {(c)};
    \end{tikzpicture}
    \begin{tikzpicture}
    \draw (0,0) node[inner sep=0]{\includegraphics[width=0.45\linewidth]{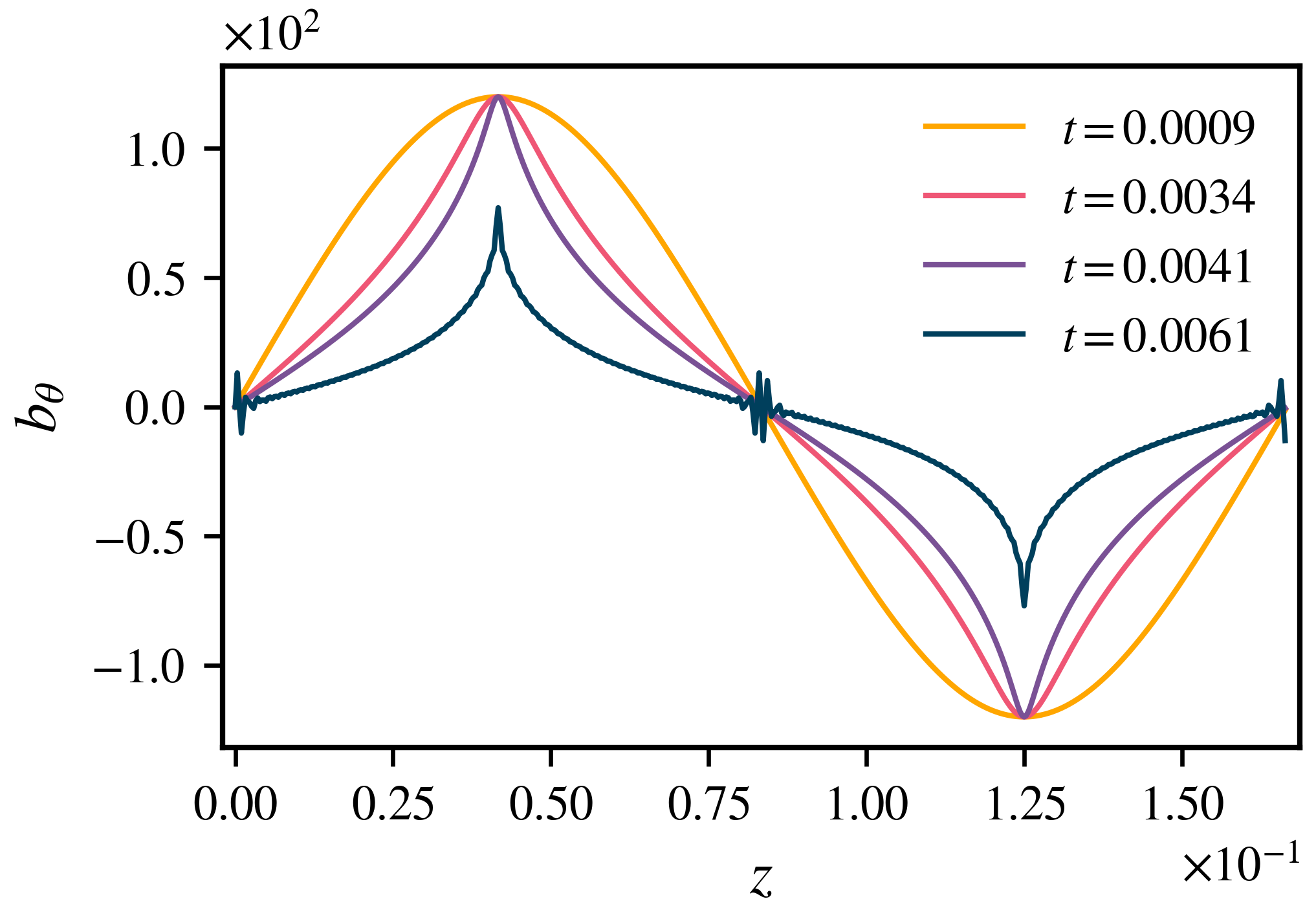}};
    \draw (-1.9,-1.2) node {(d)};
    \end{tikzpicture}    
    \begin{tikzpicture}
    \draw (0,0) node[inner sep=0]{\includegraphics[width=0.45\linewidth]{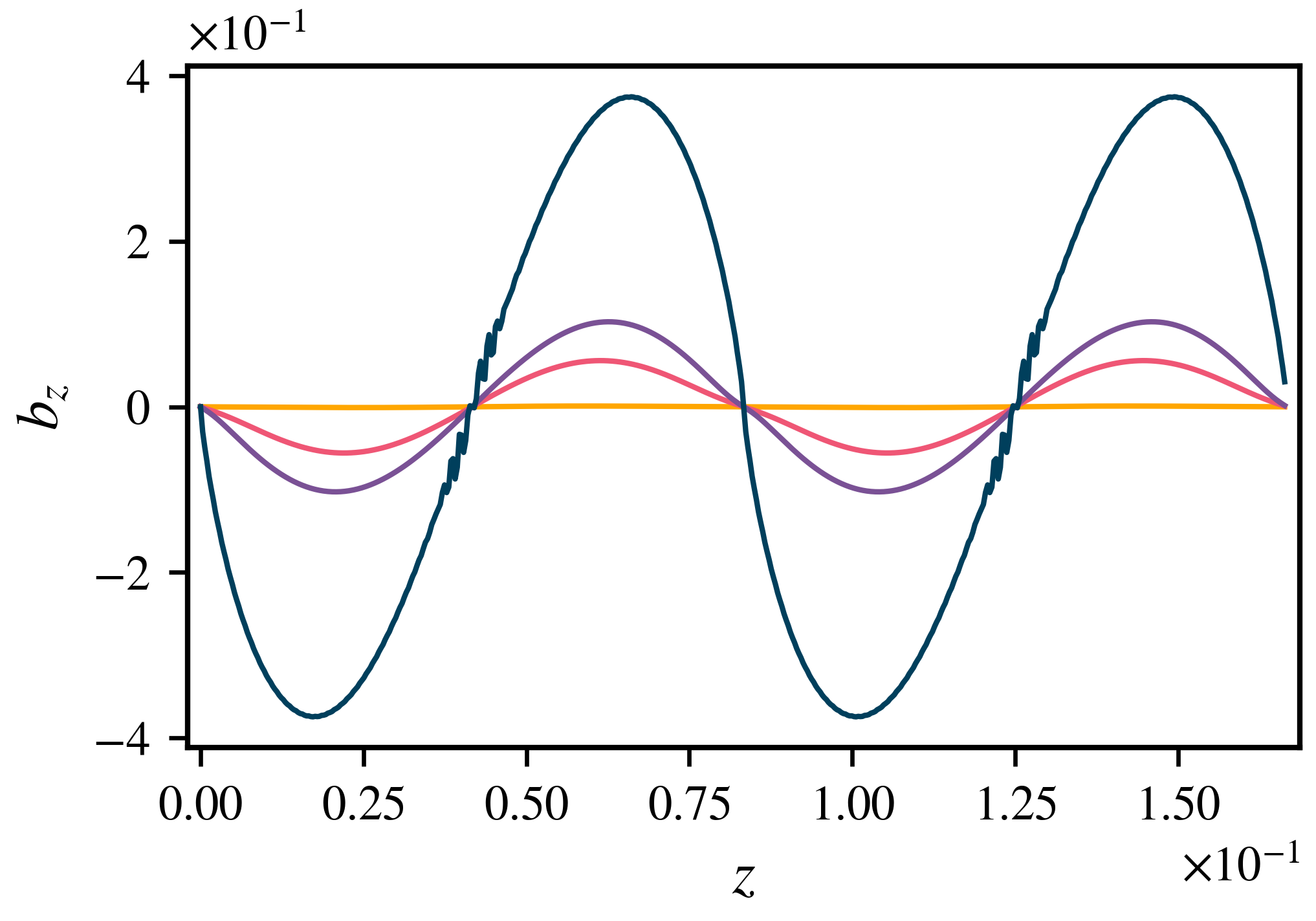}};
    \draw (-1.9,-1.2) node {(e)};
    \end{tikzpicture}
    \begin{tikzpicture}
    \draw (0,0) node[inner sep=0]{\includegraphics[width=0.45\linewidth]{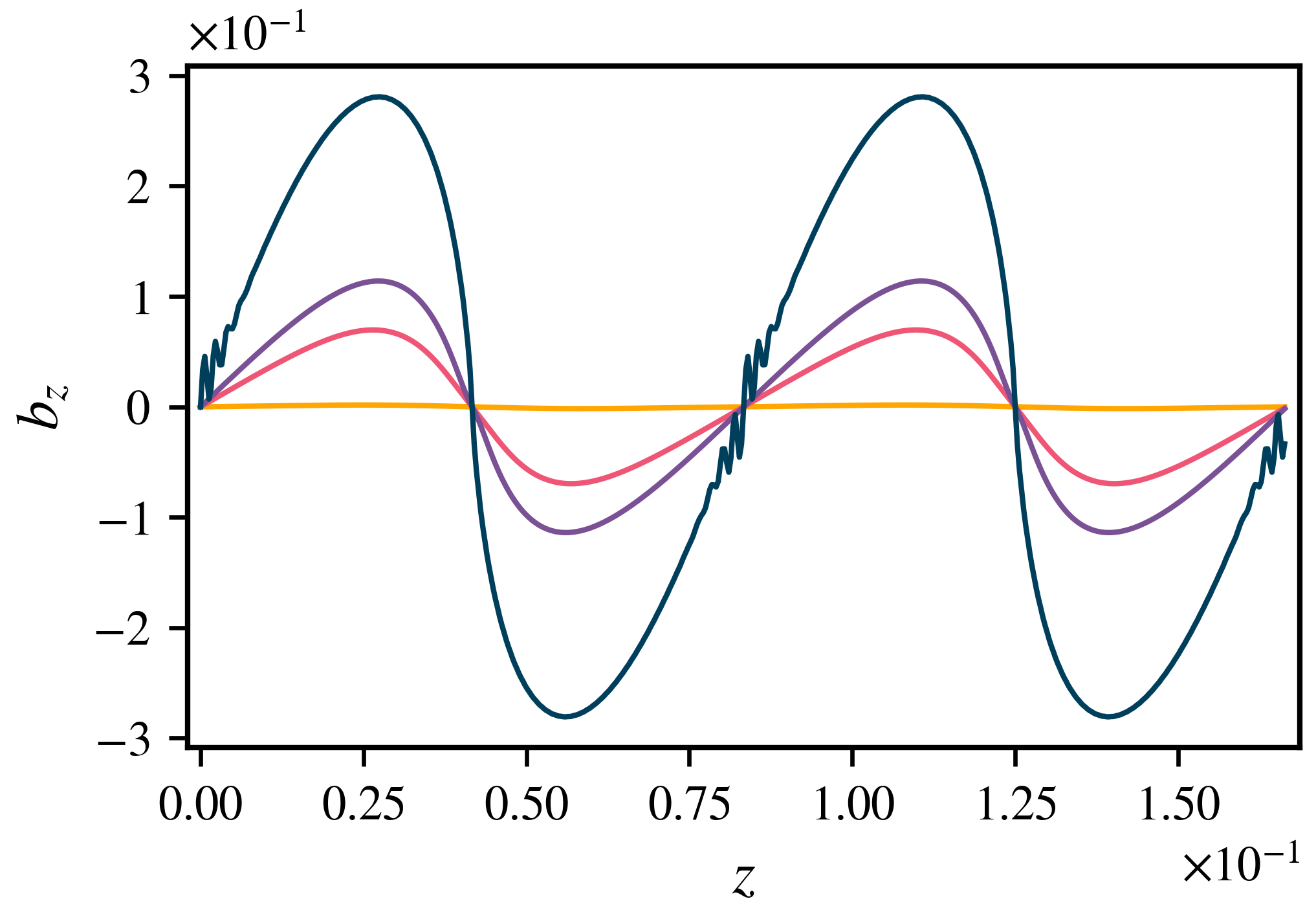}};
    \draw (-1.9,-1.2) node {(f)};
    \end{tikzpicture}    
    \caption{Plot versus $z$ of (a,b) $u^{z}$, (c,d) $b^{\theta}$ and (e,f) $b^z$ at $r=1$, superposed for various times $t$ listed in the legend. The flows have been initiated with (a,c,e) $\mathcal{C}=0.80$ and (b,d,f) $\mathcal{C}=1.20$. 
    Here, we use a resolution of $(N_r=256, N_z=512)$. At later times, we see the development of localized oscillatory structures called \textit{tygers} in these fields. }
    \label{fig:extra_tygers}
\end{figure}

\bibliography{references}